\pdfoutput=1
\documentclass[usenatbib]{mn2e}
\usepackage{hyperref}
\usepackage{apjfonts}
\usepackage{amssymb}
\usepackage{bm}
\usepackage{amsmath}
\usepackage{ctable}
\usepackage{url}
\usepackage{breakurl}
\usepackage{fixltx2e} %% forces figure order to remain fixed (otherwise figure*'s can move out-of-order)

\newcommand{\apj}{Astrophys. J.}
\newcommand{\apjl}{Astrophys. J. Lett.}

\newcommand{\mnras}{Mon. Not. R. Astron. Soc.}

\newcommand{\araa}{Ann. Rev. Astron. Astrophys.}
\newcommand{\aap}{Astron. Astrophys.}
\newcommand{\prl}{Phys. Rev. Lett.}

\newcommand{\pre}{Phys. Rev. E}
\newcommand{\pla}{Phys. Lett. A}

\newcommand{\acknowledgments}{\begin{small}\section*{Acknowledgments}\end{small}}
\newcommand\altaffilmark[1]{$^{#1}$}
\newcommand\altaffiltext[1]{$^{#1}$}
\title[Density in supersonic turbulence]{The distribution of density in supersonic turbulence\vspace{-0.5cm}}

\author[Squire et al.]{
\parbox[t]{\textwidth}{ 
	Jonathan Squire\altaffilmark{1,2} \&\ Philip F. Hopkins\altaffilmark{1}
} 
\vspace*{6pt} \\
\altaffiltext{1}{TAPIR, Mailcode 350-17, California Institute of Technology, Pasadena, CA 91125, USA} \\
\altaffiltext{2}{Walter Burke Institute for Theoretical Physics, Pasadena, CA 91125, USA\vspace{-0.3cm}}
}

\date{Submitted to MNRAS, ?, 2016\vspace{-0.6cm}}
\begin{document}
\maketitle

\begin{abstract}
We propose a model for the density statistics in supersonic turbulence, which play a crucial role in star-formation  and the physics of the interstellar medium (ISM). Motivated by 
 [Hopkins, MNRAS, {\bf 430}, 1880 (2013)], the model considers the density to be arranged into
 a collection of strong shocks of width $\sim\! \mathcal{M}^{-2}$, where $\mathcal{M}$ is the turbulent Mach number. With 
 two physically motivated  parameters, the model predicts all density statistics for $\mathcal{M}>1$ turbulence: the density 
probability distribution and its intermittency (deviation from log-normality), the density variance--Mach number relation, power spectra, and structure functions. 
For the proposed model parameters, reasonable agreement is seen between model predictions and numerical simulations, albeit within the large uncertainties associated with
current simulation results. More generally, the model could provide a useful framework for more detailed analysis of future 
simulations and observational data. Due to the simple physical motivations for the model in terms of shocks, it is straightforward to 
generalize to more complex physical processes, which will be helpful in future more detailed applications to the ISM. We see good qualitative agreement between such extensions and recent simulations of 
non-isothermal turbulence.
\end{abstract}

\begin{keywords}
star formation: general --- turbulence --- shock waves --- ISM: kinematics and dynamics
\vspace{-1.0cm}
\end{keywords}

\vspace{-1.1cm}

%%%%%%%%%%%%%%%%%%%%%%%%%%%%%%
\section{Introduction}

A detailed knowledge of gas  statistics in the  interstellar medium (ISM) is pivotal for theories of star formation and the 
stellar initial mass function (see, for example, \citealt{Krumholz:2005go,Federrath:2012em,Hopkins:2013jf,Hennebelle:2013fj}
and references therein).
The difficulty in understanding these statistics arises because 
much of the ISM is in a state of \emph{supersonic turbulence}---a highly 
chaotic tangle of interacting shocks with structure spanning  an
enormous range in scale. 
However, despite a wide range of work on the 
subject, enabled in large part by the explosive growth in computational power (e.g., \citealt{Kritsuk:2007gn,Lemaster:2008kb,Molina:2012iv,Federrath:2013gu,Pan:2015wk} and references therein), much less is known about the statistical properties of supersonic turbulence in comparison to its subsonic cousin \citep{Federrath:2013gu,Pan:2015wk}. For example, 
despite some promising results (e.g., \citealt{Boldyrev:2002uh,Aluie:2011ie,Banerjee:2013hs}), we currently lack a well-accepted theory for the velocity power spectrum, similar to the standard Kolmogorov phenomenology for subsonic turbulence \citep{1941DoSSR..30..301K}.
Further, an important aspect of supersonic turbulence theory, which  is much less relevant 
in subsonic turbulence, is the density statistics. These are crucial for  star-formation applications and directly  observable in the ISM.
However, although there are certain well-established results---most importantly the \emph{density variance--Mach number relation}: that the density distribution is approximately log-normal with a variance that 
increases with Mach number \citep{Passot:1998cr,Price:2011ha,Padoan:2011jd,Molina:2012iv,Federrath:2015gc,Pan:2015wk}---we  lack 
detailed understanding of many important issues. For example, the power spectrum of density and its variation with Mach number is not well understood  \citep{Kim:2005fq,Kritsuk:2007gn,Kowal:2007gz,Konstandin:2015kv}. Further, 
an important limitation of the density variance-Mach number relation is that the
density can be quite \emph{intermittent}, viz., it is not distributed log-normally. This behavior  manifests itself in a significant 
negative skewness in the probability density function (PDF) \citep{Federrath:2008ey,Price:2010ds,Schmidt:2009dr,Konstandin:2012ei,Hopkins:2013hn}, enhancing the probability 
of low density regions while decreasing the probability of high density regions.
The purpose of this work is to propose and examine a simple phenomenological model for the turbulent density field  that encompasses all such statistics: the power spectrum, the density PDF and intermittency, and how these vary with turbulent Mach number.

It is worth elaborating on the density intermittency mentioned in the previous paragraph: is this  important, or simply a formal nuisance to occupy the idle theorist?
For some applications, in particular those that depend on events that occur regularly (i.e., 
with high probability), the answer is probably that intermittency is not important: so long as the density PDF is 
approximately log-normal near its peak, high-probability events will be well-characterized 
purely by the variance. However, many physically interesting properties that one might wish 
to derive from turbulent PDFs involve rare events, in particular those involving  high-density regions.
As a simple example, if we were interested in regions with a  factor $\sim 100 $ enhancement in density over
the mean  at a turbulent Mach number  $\mathcal{M}\sim 15$ (e.g., to push the density above the Jean's density and enable gravitational collapse), the probability of finding such a region could easily be over-estimated by $1$ to $ 3$ orders of magnitude by a  log-normal model, compared to a more realistic intermittent 
model with the same variance.
Gaining a better understanding of the density PDF can also be motivated by other more 
fundamental interests. For instance, averaged over scale $l$, the PDF of the density field \emph{cannot} be log-normal (note that $l$ could be the grid scale here): by predicting
a nonzero probability of finding densities greater than the ratio between the box volume and $l^{3}$, a 
log-normal PDF violates mass conservation
(see also \citealt{Hopkins:2013hn}; hereafter \defcitealias{Hopkins:2013hn}{H13}\citetalias{Hopkins:2013hn}).
Further, we shall find that the process of understanding 
the origin of the density intermittency using the properties of shocks leads to various other insights;
for example, the origins of the qualitatively different PDF shapes seen in non-isothermal or magnetohydrodynamic (MHD) turbulence.

This work involves an extension of the supersonic density PDF model proposed by \citetalias{Hopkins:2013hn} to multi-point statistics.
Specifically, this involves specifying the properties of the PDF \emph{as a function of scale}, viz.,
after averaging the density field over scale $l$ (which is some fraction of the box scale $L$) what is its PDF?
Such a model completely specifies the statistical properties of the turbulent density field and contains significantly more information than the 1-point PDF that is usually calculated from numerical studies. For example, it allows the prediction of
the density variance-Mach number relation, a similar relation for the
intermittency, density power spectra, and structure functions.
Further, these predictions are made with just one or two physically motivated free parameters, that are
 independent of Mach number $\mathcal{M}$ once this is large.
Extending the discussions in \citetalias{Hopkins:2013hn} and based on She-Leveque intermittency models
that have been successful for subsonic turbulence \citep{She:1994fn,She:1995vi,Dubrulle:1994ta,Castaing:1996ew,He:1998du}, the model relies
on a simple physical picture in which the density field is made up of a series 
of shocks covering a wide range of scales down to where the turbulence becomes subsonic. 
Starting from the box scale, at which the density field is simply constant, the
shocks add density variance in the form of discrete multiplicative events, where 
the size of individual events is  related to the physical properties of a shock.
This predicts a strong relationship between the size of individual events---which controls the intermittency---
and the variance of the density PDF, governed by the range of scales (i.e., $L/l$) and the Mach number.
Comparison to various well-known results and trends, as well as our own numerical simulations, 
illustrates reasonable agreement across a range in Mach numbers. 
In addition, given the simple physical reasoning used to derive model parameters from
 isothermal shock properties, we
extend the model to more complex and realistic physics---for example, a non-isothermal 
gas equation of state, or MHD---explaining various aspects of the density PDFs and why these are different from isothermal turbulence. 

The remainder of the paper is organized as follows. In Sec.~\ref{sec: general considerations}, we outline a few general considerations that will be used to motivate various choices in our model. In Sec~\ref{sec:model} we explain the model, in particular 
how the mathematical structure of a (compound) log-Poisson random process can be related to 
the properties of individual shocks. We take particular
care here to outline the choices necessary for various parameters, and
how these may be motivated or phenomenologically derived from physical properties of 
the turbulence. We then outline a variety of predictions of the model---including 
the density variance--Mach number relation, intermittency predictions, and power spectra---and compare these to numerical simulations. This comparison involves both previous results and
a variety of new simulations using the Lagrangian Meshless-Finite-Mass method in the GIZMO code \citep{Hopkins:2015bk}, 
which we use to directly compute the density PDF as a function of scale. We also measure the physical size of shock structures, which forms an important part of our argument, in App.~\ref{app:shock sizes}.
Overall, model predictions seem to match with 
numerical results up to numerical uncertainties, although more detailed  
comparisons will be necessary to  understand its successes and failures more completely.
We finish with an extension  to 
non-isothermal turbulence (illustrating reasonable qualitative agreement with 
the simulations of \citealt{Federrath:2015gc}) and a discussion of MHD, before concluding by 
reiterating the model's main predictions.

\section{General considerations}\label{sec: general considerations}

Before continuing, it seems worth enumerating several general points about 
supersonic turbulence. While some of these are well known,
given that each plays some crucial role in the derivation of our model,
it is helpful to clearly explicate these ideas early on in our discussion.
\begin{enumerate}
\item {Supersonic turbulence is not scale invariant, except in the infinite Mach number limit. 
This arises because of the importance of the sonic scale $l_{\mathrm{sonic}}$, which is the scale at which
$v_{l}=v_{|\bm{l}|}= \langle | v(\bm{x}+\bm{l}) - v(\bm{x})|\rangle \sim c_{s}$; i.e., the scale at which the turbulence 
becomes subsonic.  For modest Mach numbers common in nature or numerical experiments, the scale separation between $l_{\mathrm{sonic}}$ and the driving scale is also  modest, challenging the relevance of the concept of a supersonic ``inertial range.''  This 
 feature necessarily leads to some important differences in the theoretical treatment of supersonic
 turbulence in comparison to subsonic turbulence.}
 \item{Since density is effectively defined with reference to a volume (it is the mass per unit volume), 
the density itself, and its PDF, are naturally defined with reference to an averaging scale. This density PDF as a function
of scale encodes a wide variety of useful statistical information about the density field.
By only ever studying the PDF with respect to some arbitrary scale (usually the grid scale in 
numerical simulations), one risks missing important trends or effects, particularly considering that the
sonic scale is often close to the grid scale for Mach numbers of $\sim\! 10\rightarrow 20$ at currently 
available numerical resolutions (see point 1). As an example, in the numerical simulations in this work, 
we find that the density PDF becomes significantly \emph{less} intermittent below $l_{\mathrm{sonic}}$, presumably
because the density field on subsonic scales involves nearly Gaussian fluctuations \citep{Federrath:2010ef,Konstandin:2012ei}.
The density averaged over scale is also relevant for applications, being an important quantity for studying 
gravitational collapse. For example, a region is of (linear) size $l$ is unstable to collapse 
if $l$ is larger than the Jean's length 
\begin{equation}
\lambda_J \sim \frac{c_s}{\sqrt{G \rho_l} }. 
\end{equation}
Here $\rho_l$ is the average density over $l$, so to understand the statistics of gravitational collapse in a turbulent cloud, we  require an understanding of the density PDF as a function of scale. }
\item{A PDF that appears more log-normal does not necessarily  imply that the underlying statistics are closer
to Gaussian. Instead, the apparent Gaussianity of the PDF may be a result of the suppression of 
low densities compared to the isothermal case.
This point is important for MHD and non-isothermal turbulence.   }
\item{The assumption that the shock width is equal to the sonic scale---while 
 useful as a phenomenological tool for deriving density variance--Mach number relations \citep{Price:2011ha,Padoan:2011jd,Molina:2012iv,Federrath:2015gc}---is inconsistent with  log-normal density statistics. In particular, 
for isothermal shocks with a density contrast $\sim \mathcal{M}^{2}$, such a model involves \emph{all} of the mass being concentrated in a single shock, which 
leads to an unphysically intermittent density distribution.  In our model, we take the shock width to be some 
small fixed fraction $\kappa$ of $l_{\mathrm{sonic}}$, and it will transpire that the $\kappa$ parameter controls the intermittency. The
 success of density variance estimates using $l_{\mathrm{sonic}}$ as the shock width may then be related to $\kappa $ being
approximately universal, even in more complex physical situations (e.g., MHD).}
\item The infinite-Mach-number limit is not equivalent to Burger's turbulence (where the pressure term is neglected in the Navier-Stokes equations). This is because there are always some regions, no matter how large $\mathcal{M}$, where the pressure forces becomes important; see \citet{Passot:1998cr}.
\end{enumerate}

%%%%%%%%%%%%%%%%%%%%%%%%%%%%%%
\section{Model description}\label{sec:model}

In this section we describe the mathematical structure and physical motivation for the density model. 
We shall denote  the 
scale over which the density is averaged as $l$, the box scale as $L$, the PDF 
of the density averaged over scale $l$ as $\mathcal{P}_{l}(\rho)$, and assume that
 the volume average of $\rho$ over the whole box is $\langle \rho \rangle= M/L^3=1$. 
Start by considering the density averaged over the
scale of the box $l=L$, which by definition is $ \rho =1$ with the PDF $\mathcal{P}_{L}(\rho)=\delta(\rho-1)$. 
The model then provides a description of the  density PDF averaged over successively smaller subvolumes of the box, $\mathcal{P}_{l}(\rho)$, a full knowledge of which  effectively provides a full description of
the statistical state of the turbulent density field. 

\begin{figure}
\begin{center}
\includegraphics[width=0.7\columnwidth]{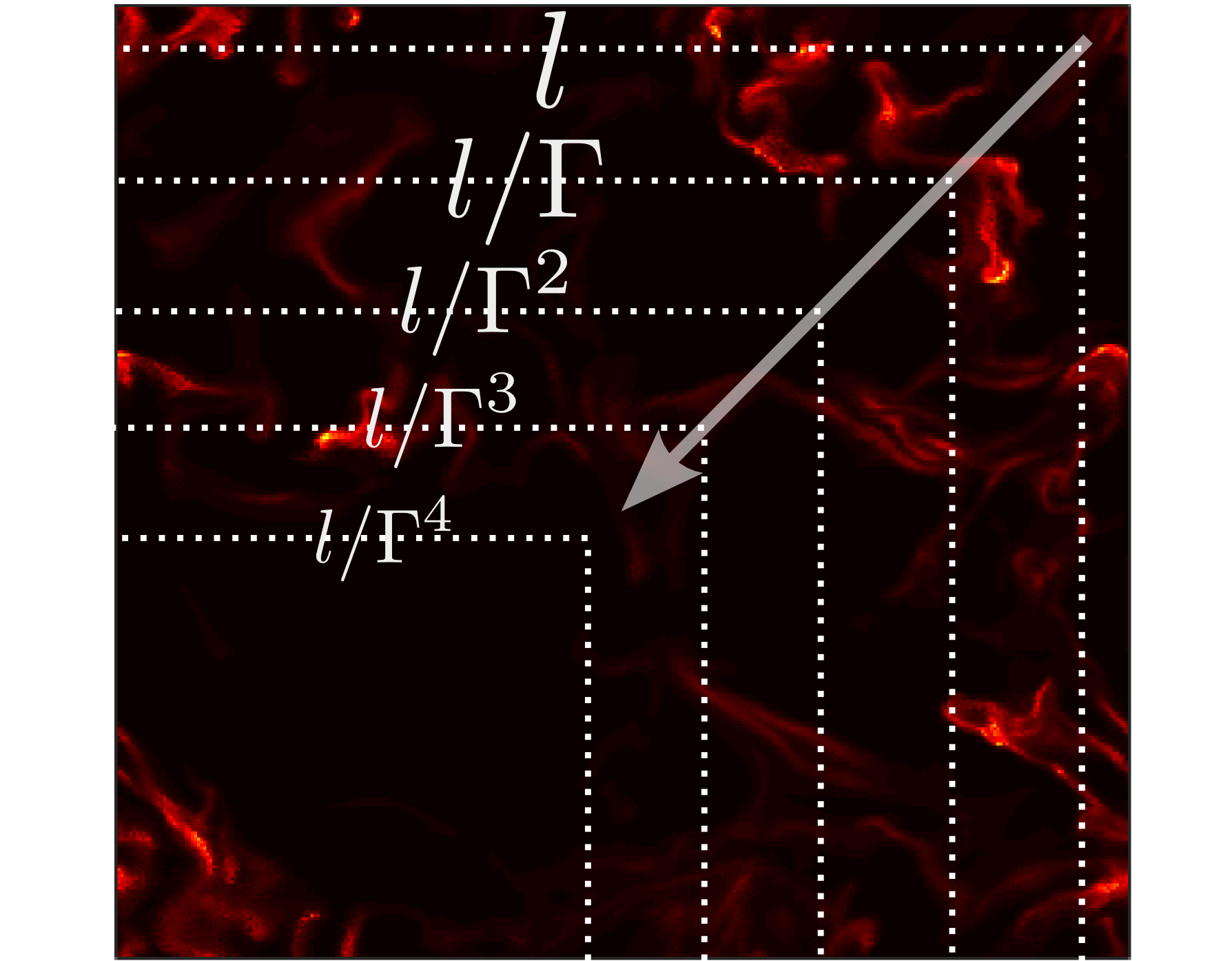}
\caption{Cartoon of how the average density on scale $l$ can change as $l$ decreases,
projected into the 2-D plane.
The brightness illustrates the density field, which is dominated by thin shock structures with a large proportion of the
total mass. In the 
hypothetical step in scale from $l$ to $l/\Gamma$ the difference in volume (between
the squares) covers a very high density shock region (upper right), implying the average density at scale $l/\Gamma$
is \emph{less} than that at $l$, since a significant proportion of the mass is lost in the jump. In contrast, in the jump from $l/\Gamma$ to $l/\Gamma^{2}$,
which entails the same proportional volume change, there is not a significant change in density because
all high density regions remain inside the new volume at scale $l/\Gamma^{2}$.
}
\label{fig:lambda jumps}
\end{center}
\end{figure}

This process is described mathematically by a series of $N$ steps,  
each of which decrease the volume over which $\rho $ is averaged by a factor $\Gamma>1$, such that $l^{3} = L^{3}/\Gamma^{N}$.    The basic idea is that as one makes a jump down in scale from $l$ to $l/\Gamma$, there is some probability, scaling with $\Gamma-1$, of an ``event'' that changes the density by $\delta \rho$. This event will be related to a shock structure that was previously 
in the larger volume (at scale $l$) being lost from the smaller volume (at scale $l/\Gamma$), thus
causing a decrease in density in the volume being considered. 
This shock is related to velocity and density structures that vary over scale $l$, since
smaller structures cannot contribute significantly due to their smaller size, while there 
are too few larger structures (since these vary over scales $\gg l$).
The size of this ``jump'' decrease in density $\delta \rho$
is related to the proportion of mass that resides in the shock that is lost in the step, and is itself a random variable. Assuming that most of the mass in the system is tied up in shocks, 
steps with no event will cause the average density to increase slightly,  because the density 
in the volume that is lost in the step is less than the average density in the system.
We graphically illustrate such a process in Fig.~\ref{fig:lambda jumps}, showing how in a volume filled 
with high density shocks, some steps will cause a large decrease in the mass enclosed by the new volume (i.e., an event), while other steps will not (no event).

\subsection{Compound-log-Poisson cascades}

The process we have just described is known mathematically as a \emph{compound-log-Poisson} random process. Such processes are well characterized and have been particularly successful
in phenomenological turbulence intermittency models for  velocity structure functions
\citep{She:1994fn,Dubrulle:1994ta,Castaing:1996ew,He:1998du,Boldyrev:2002uh,Mallet:2016}. 
The process occurs in log space because we consider multiplicative changes to the volume and density (which are additive in log space), while 
it is a Poisson process because the probability of an event is proportional to the size of the jump in volume (with infinitesimally small jumps being considered). 
If the size of each density jump, $\ln \rho \rightarrow \ln\rho - \delta \ln \rho$, is fixed 
(i.e., if $\delta \ln \rho$ is a number), then the resulting distribution will be log-Poisson,
\begin{equation}
\mathcal{P}(\ln\rho) = \frac{\lambda^{-\ln\rho + \Upsilon}e^{-\lambda}}{(-\ln\rho + \Upsilon)!},\label{eq: standard logPoisson}
\end{equation}
where the $\Upsilon$ accounts for a mean shift in the distribution (required to keep $\langle \rho\rangle=1$; see below). 
However, the distribution \eqref{eq: standard logPoisson} has the unphysical property of being discrete, since the jump sizes are discrete, and is thus somewhat inconvenient as a model for density.
This issue may be circumvented by postulating that the size of each jump $\delta \ln \rho$ is itself
a random variable, with probability distribution $\delta \ln \rho \sim \mathcal{P}_{\delta}(\delta \ln \rho)$. This leads to a ``compound-log-Poisson'' distribution, where  ``compound'' refers to the idea that the 
distribution is formed as a random process of a random variable.
Following \citetalias{Hopkins:2013hn}, we take the jump sizes to be distributed exponentially \citep{Castaing:1996ew},
\begin{equation}
\mathcal{P}_{\delta}(\delta \ln \rho) = \begin{cases}
T^{-1}\exp [(\delta \ln \rho - \epsilon) /T] & \delta \ln \rho - \epsilon<0\\
0 & \delta \ln \rho - \epsilon>0\\
\end{cases},\label{eq:step PDF}\end{equation}
where $T$ is the mean jump size and $\epsilon$ is a constant that is used to ensure 
$\langle \rho \rangle=1$; see Fig.~\ref{fig:stepPDF}.
 This leads  to a  convenient 
form for the PDF that matches  density PDFs measured from simulations
remarkably well \citepalias{Hopkins:2013hn}.\footnote{The choice of an exponential distribution for $\mathcal{P}_{\delta}(\delta \ln \rho)$ can 
be motivated as the only choice other than a delta-function distribution (i.e., a standard log-Poisson process)
that leads to a single fractal dimension for the most singular structures  \citep{He:1998du}. 
However, since the system we model is not scale invariant anyway, our motivation for this choice 
is primarily simplicity---the PDF may be written in a simple closed form without unphysical discrete jumps---and other choices give  similar results.}

% The approach we take here is related to these models, but we shall place more emphasis on constraints 
% arising from mass conservation (which should be important for density but not necessarily velocity), as opposed
% to the dimensionality of the most singular structures. As some motivation for this change in perspective, we note that the success of 
% these intermittency models is likely related to fact that a (compound) log-Poisson process is scale invariant;
% however, unlike subsonic turbulence, supersonic turbulence at moderate Mach number is explicitly \emph{not} scale invariant,
% given the importance of the sonic scale where $\mathcal{M}=1$.

We now derive the PDF of $\ln \rho$, $\mathcal{P}_{l}(\ln \rho)$. that arises from this process.
Assuming that the parameter $T$ does not depend on scale, $\mathcal{P}_{l}(\ln \rho)$ is the convolution of 
$n$ $\mathcal{P}_{\delta}$ distributions shifted by the total number of steps $\Upsilon = N \epsilon$,
\begin{equation}
\mathcal{P}_{l}(\ln\rho) = \mathcal{P}_{\delta}^{\otimes n} = \mathrm{Gamma}(- \ln \rho +\Upsilon;n,T),
\end{equation}
where $\cdot^{\otimes n}$ denotes the convolution power and $\mathrm{Gamma}(x;n,T) = x^{n-1}e^{-x/T} T^{-n}\Gamma(n)^{-1}$ is the Gamma distribution.
As one takes the limit $N\rightarrow \infty$, $\Gamma \rightarrow 1$ with $N(\Gamma-1) = \lambda$, the number of events $n$  is itself is a Poisson-distributed random variable with mean $\lambda$ (this 
will depend on scale $l$ and is specified below), so the full PDF for $\ln \rho$ is simply a sum
of the PDFs for a given $n$, weighted by the probability that such an $n$ occurs $n\sim \lambda^{n}e^{-\lambda}/n!$.
Putting this together, one obtains
\begin{align}
\ln \rho \sim \mathcal{P}_{l}(\ln \rho) & = \sum_{n=0}^{\infty}\frac{\lambda^{n}e^{-\lambda}}{n!} \frac{u^{n-1}}{T\Gamma(n)}\exp(-u)\nonumber \\
&= T^{-1}\sqrt{\frac{\lambda}{u}}I_{1}(2\sqrt{u \lambda})\exp[-(\lambda+u)],\label{eq:Phils}
\end{align}
where $u\equiv (-\ln \rho + \Upsilon)/T$, $I_1(x)$ is the first-order modified Bessel function of the first kind, and $ \mathcal{P}_{l}(\ln \rho)$ is nonzero only for $u>0$. 
We may then fix $\Upsilon= N\epsilon = \lambda T(1+T)^{-1}$ using the constraint $\langle e^{\ln \rho}\rangle=\langle\rho\rangle=1$, which leads to the density PDF proposed in \citetalias{Hopkins:2013hn}. The volume-weighted variance is $S_{\ln \rho,V} = 2\lambda T^{2}$, while $T$---the mean size of the jumps---is an intermittency parameter that skews the distribution towards higher probability at $\ln \rho <0$. 
In line with our intuition, small numbers of large density jumps (e.g., large shocks)
lead to highly intermittent distributions, while a large number of small
jumps leads to distributions that are very close to log-normal. The mass-weighted variance is $S_{\ln \rho,M} = 2\lambda T^{2}(1+T)^{-3}$, so $S_{\ln \rho,M} = S_{\ln \rho,V} $ in the $T=0$ (log-normal) limit as expected \citepalias{Hopkins:2013hn}. Also note
that, unlike a log-normal PDF, this distribution has a maximum value $\ln \rho = \Upsilon$ above which $\mathcal{P}_{l}(\ln \rho)=0$. This property is 
entirely physical: the probability of encountering $\rho> \langle\rho\rangle(L/l)^{3}$ is identically zero, since there
is not enough mass in the system.

To relate  the mathematical model \eqref{eq:Phils} to the physical 
properties of supersonic turbulence, we require two extra pieces of information: (i) how the average number of events $\lambda$ relates to the 
physical scale $l$, and (ii) how the parameter $T$ (the size of an event) 
relates to the physical properties of a shock structure. We tackle these issues in the next two sections.

\subsection{Variation with scale and mass conservation}

 \begin{figure}
\begin{center}
\includegraphics[width=0.8\columnwidth]{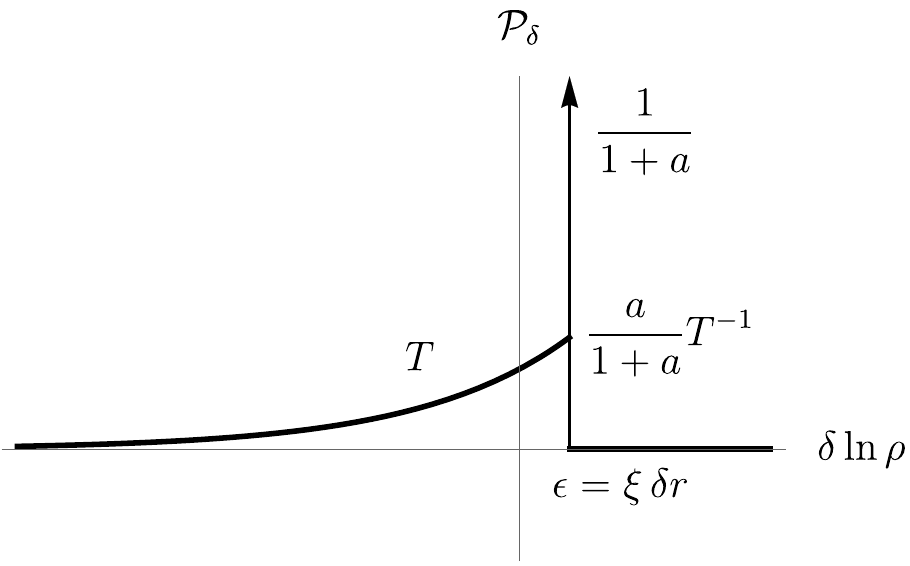}
\caption{The PDF for a single step in the cascade, which is $a(1+a)^{-1} \mathcal{P}_{\delta}(x) + (1+a)^{-1} \delta(x)$,
where $a$ is a small parameter proportional to the size of the step in volume $\delta r = \ln(\Gamma -1)$. 
The maximum $\epsilon$ is chosen as $\epsilon = \xi\delta r$, which   correctly captures
the true maximum of $\rho$ when $\xi =3$, and otherwise describes how the 
maximum density changes with scale.
When coupled with the constraint $\langle \rho \rangle=1$, $\xi$ sets the proportionality between 
$a$ and $\delta r $ as $a = \xi \delta r (1+T)T^{-1}$. }
\label{fig:stepPDF}
\end{center}
\end{figure}

In this section, we relate the average number  of events $\lambda$  to the physical scale $l$.
To explore this, it is helpful to consider the PDF of $\delta \ln \rho$ 
for a single step, which is illustrated in Fig.~\ref{fig:stepPDF}. This is a mixture distribution of $\mathcal{P}_{\delta}$ with probability $a/(1+a)$ (there is an event), and a (shifted) $\delta$-function distribution with probability $1/(1+a)$ (there is no event). The parameter  $a$ thus determines the
mean number of events over a given number of steps, implying the relation between $\lambda$ and $l$ is controlled by the proportionality constant 
between $a$ and the (log) differential change in scale between $l$ and $l/\Gamma$, $\delta r = \ln\Gamma \approx \Gamma -1$. 

The first constraint we can apply to the single-step PDF is that the total mass must conserved, $\langle \delta \rho \rangle_{\delta} = \langle e^{\delta \ln \rho +\epsilon} \rangle_{\delta} =1$ (where $\langle\cdot \rangle_{\delta}$
indicates the mean over the single-step PDF). This leads 
to the constraint \begin{equation}
a = \epsilon \frac{1+T}{T}.
\end{equation}
The second mass-conservation-related constraint we apply is that the \emph{maximum} density   scales 
as $(L/l)^{\xi}$; i.e., $\epsilon = \xi \delta r$, or 
\begin{equation}
\lambda_l = Na = \xi  \frac{1+T}{T} \ln\left(\frac{L}{l}\right).\label{eq: general lambda form}
\end{equation}
This constraint  simply states that the increase in density  
when there is no event is proportional to some power of the change in linear dimension. If $\xi=3$,
this implies that \emph{all} of the mass  from the old volume is 
contained into the new volume when there is no event, which will give a density PDF with the true (physical) density maximum $\rho_{\mathrm{max}}=(L/l)^3$. 
However, we find empirically (see Sec.~\ref{sec:Numerics})
that $\epsilon = 3 \delta r$ predicts a PDF that is insufficiently 
intermittent, and that $\epsilon \approx 1.5\delta r$  provides a closer fit to data\footnote{Although it is difficult  motivate a particular value for  $\xi$, a value $\xi<3$ fits our intuitive picture of a collection of shocks. In particular,  even in the case when all of the mass
is contained in infinitely thin shocks, most  of these will be at some angle to the changing volume and thus a ``no event'' step will involve losing a small part of this mass. For example, if such shocks on average intersect one dimension of the changing volume, this would imply $\xi = 3-1=2$, while if they intersect two it would imply $\xi = 3-2=1$. Thus, $\xi$ is related to the density in unshocked regions, the dimensionality of the shocks, and their distribution in angle.} (for the model of $T$ proposed in Sec.~\ref{subsec:T discussion} below).  We thus consider $\xi$ to be a parameter of our model, which should be universal for all turbulence with $\mathcal{M}\gg 1$. 
Note that a deviation from this general form (i.e., if $\epsilon$ was not taken proportional to $\delta r$ but chosen through some other method) would imply that the geometrical properties and distribution of the shocks was a function of scale. This would be inconsistent with the existence of an inertial range in the $\mathcal{M}\gg 1$ limit, but could in principle occur for $\mathcal{M}\sim 1$.

\subsection{Relating $T$ to the size of a shock}\label{subsec:T discussion}

In this section, we relate the intermittency parameter $T$ to the 
physical properties of individual shocks (see also \citetalias{Hopkins:2013hn} appendix). 
In the 
moderate-$\mathcal{M}$ regime, which is both physically relevant and most commonly probed by current numerical simulations, the system is not scale invariant and $T$ must be an increasing function of $\mathcal{M}$, eventually asymptoting to some value $T\lesssim 1$.\footnote{Note that  $T>1$ implies most of the mass in the system is contained in a single shocked structure, which might cause one to question whether the system is truly turbulent.} 
Here we propose a simple physically motivated model, which gives a good match to simulations (see Sec.~\ref{sec:Numerics}), is easily extendable to other equations of state (see Sec.~\ref{sec: Extensions}) and is based on well-accepted ideas put forth in  previous literature \citep{Passot:1998cr,Price:2011ha,Padoan:2011jd,Molina:2012iv}. However, 
a variety of other possibilities exist, some of which could prove similarly successful and also be physically motivated. This is discussed further in Sec.~\ref{sub:model choices}.

 ``Shocks'' will be taken as regions
that are extended across the volume being considered (i.e.,  of size $l$) in two-dimensions, 
and be of  small but finite extent in the other dimension \citep{Federrath:2008ey}. Concentrating 
for now on the simplest isothermal equation of state, the Rankine-Hugoniot 
conditions for mass and momentum conservation of $\rho$ and $v$ on either side of a shock are 
\begin{equation}
\rho_{1}v_{c1}=\rho_{2}v_{c2},\quad \rho_{1}(v_{c1}^{2}+c_{s}^{2})=\rho_{2}(v_{c2}^{2}+c_{s}^{2}),\label{eq:RK conds}\end{equation}
where $v_{ci}$ is the velocity on side $i$ in the direction perpendicular to the shock. 
Solution of Eq.~\eqref{eq:RK conds} with $v_{c0} = b\mathcal{M}c_{s}$, based on some 
average Mach-number $\mathcal{M}$ and average fraction of the velocity in compressive modes $b$ \citep{Padoan:1997bh,Passot:1998cr,Federrath:2008ey},  leads to an approximate relation for the density contrast:
\begin{equation}
\frac{\rho_{1}}{\rho_{0}} \sim b^{2}\mathcal{M}^{2}.\label{eq:shock contrast}\end{equation}
As common in previous works \citep{Lemaster:2008kb,Padoan:1997bh,Passot:1998cr}, we use Eq.~\eqref{eq:shock contrast} to relate the density in a shock to that outside the shock, for each scale in the turbulence.

The important quantity controlling $T$ is the average density jump
that occurs when a shock is removed from the volume being considered. This depends on both the density contrast and the volume of shocks, which requires a measure of 
their physical width $r_{\mathrm{shock}}$.
Given that there is only one scale in the system---the sonic scale $l_{\mathrm{sonic}}$,
at which $v\sim c_{s}\,$---we effectively have only one choice: that 
the shock has width $\kappa l_{\mathrm{sonic}}$, with $\kappa <1$ some arbitrary parameter. Denoting $v_{l}$ as the approximate
velocity difference across scale $l$ and taking $v_{l}\sim (l/L)^{\zeta} \mathcal{M} c_{s}$ with $\zeta \sim 1/2$ (i.e., a velocity power spectrum $E\sim k^{-1/\zeta}$; \citealt{Federrath:2013gu}),
one obtains $l_{\mathrm{sonic}}/L\sim \mathcal{M}^{-2}$. 
Given the scaling of the density contrast Eq.~\eqref{eq:shock contrast}, we see that this prescription $r_{\mathrm{shock}}\sim \kappa l_{\mathrm{sonic}}$, states that a fixed fraction  of the mass ($\sim \kappa b^2$) is contained in individual shocks in the high-$\mathcal{M}$ limit. 

Denoting the density in the shock $\rho_{\mathrm{shock}}$, 
the density outside the shock $\rho_{\mathrm{out}}$, and the average density over both regions $\rho_{\mathrm{av}}$, we see from Eq.~\eqref{eq:step PDF} that 
\begin{equation}
T \approx \ln \frac{\rho_{\mathrm{out}}}{\rho_{\mathrm{av}}}
\end{equation}
(where we have neglected the offset $\epsilon$ since $\delta r$ may be taken to be small). 
If we then take $\rho_{\mathrm{shock}}\sim \mathcal{M}^2\rho_{\mathrm{out}}$ (absorbing $b$ into $\kappa$),\footnote{We thus expect $\kappa$ to differ between compressibly and solenoidally forced turbulence \citep{Federrath:2013gu}. Of course, $b$ and $\kappa$ are  not quite equivalent, relating to the shock density contrast and volume respectively. Nonetheless, retaining $b$ separately leads to nearly the same relation as 
Eq.~\eqref{eq:T} (with $\kappa\rightarrow b^2 \kappa$), but with unphysical ($T<0$) behavior near $\mathcal{M}=1$. In any case, $\kappa$ and $b$ are each based on  heuristic ideas, and assigning too much physical 
relevance to the details of this model is not particularly productive.} 
then from mass conservation (where $V=l^3$ is the volume),
\begin{equation}
\rho_{\mathrm{out}}(V-V_{\mathrm{shock}}) + \rho_{\mathrm{shock}}V_{\mathrm{shock}} = V \rho_{\mathrm{av}},
\end{equation}
and $V_{\mathrm{shock}}\sim \kappa l_{\mathrm{sonic}} l^2$, we obtain
\begin{equation}
T\sim \kappa (1-\mathcal{M}_l^{-2}) = \kappa \left(1- \frac{L}{l}\mathcal{M}^{-2}\right),\label{eq:T}
\end{equation}
 assuming small $\kappa$.

We thus have a simple connection between the mathematical log-Poisson cascade described above and 
the physical size  and density structure of shocks.
It is worth noting that, given the isothermal shock jump relation, making  the assumption ``shock width $\sim l_{\mathrm{sonic}}$'' (without some extra factor $\kappa <1$) is technically inconsistent with 
Gaussian statistics, although it is often used to estimate the density variance \citep{Padoan:2011jd,Molina:2012iv,Federrath:2015gc}.
Systems with a larger 
proportion of the mass in a small number of shocks will  have more intermittent statistics, since $T$ controls the deviation
from Gaussianity in Eq.~\eqref{eq:Phils}. 
In App.~\ref{app:shock sizes} we measure the shock sizes from simulation, finding reasonable agreement
with the hypothesis that they scale as some fraction of $l_{\mathrm{sonic}}$ (see Fig.~\ref{fig:shock sizes}).
Finally, we note that since we consider only density changes arising 
from shocks, we are explicitly neglecting all scales $l<l_{\mathrm{sonic}}$ where $\mathcal{M}\lesssim 1$; that is, 
the property $T(\mathcal{M}_{l}<0) = 0$  is a consequence of our focus 
on shocks as drivers of density change and is not physical. Of course, 
subsonic motions do cause variation in density, and a model for $T(\mathcal{M}<1)$ could
be added to the supersonic model presented here if so desired \citep{Federrath:2010ef}.

\subsection{The model}

 We have now completely specified the full, scale-dependent statistics for the density field. 
Assuming for the moment that $T=\kappa$ is constant for  $l>l_{\mathrm{sonic}}$ [this is
 true for $\mathcal{M}\gtrsim5$; see Eq.~\eqref{eq:T}],
 we  obtain a simple, two-parameter ($\kappa$ and $\xi$) model 
 for the density PDF as a function of scale
 \begin{gather}
 \ln \rho \sim \mathcal{P}_{l}(\ln \rho) 
 \approx T^{-1}\sqrt{\frac{\lambda}{u}}I_{1}(2\sqrt{u \lambda})\exp[-(\lambda+u)],\nonumber\\
\lambda =  \xi \left(1+\frac{1}{T}\right) \ln \left(\frac{L}{l}\right),\quad T=\kappa,\nonumber\\ u = -\ln \rho + \xi \ln \left(\frac{L}{l}\right)\quad(\mathrm{with }\:u>0),\label{eq:model}
\end{gather}
for $l>l_{\mathrm{sonic}} = L \mathcal{M}^{-2}$. For $l<l_{\mathrm{sonic}}$, we take $ \mathcal{P}_{l}(\ln \rho) =  \mathcal{P}_{l_{\mathrm{sonic}}}(\ln \rho) $; that is, 
we neglect the subsonic contribution, which is minor for $\mathcal{M}\gg 1$ \citep{Federrath:2010ef}.
Note that the Mach number dependence is implicit in Eq.~\eqref{eq:model} through the dependence 
on $l_{\mathrm{sonic}}$, since higher $\mathcal{M}$ will lead to larger $L/l$ with $l>l_{\mathrm{sonic}}$, and thus larger $\lambda$.
The difference between compressible and solenoidal large-scale motions (e.g., due to forcing; \citealt{Schmidt:2009dr,Federrath:2013gu}) is absorbed 
into the parameter $\kappa$ that controls the width and density contrast of individual shocks. 
 
 Since $T$ is not truly constant, the functional form of the  PDF will differ 
 somewhat from Eq.~\eqref{eq:model}. In fact, even the \citetalias{Hopkins:2013hn}
form [Eq.~\eqref{eq:Phils}]  is not
 produced by the model with scale-dependent $T$, since the PDF of each step varies with $l$.
 While  in principle one needs to take the convolution
 of a series of distributions $\mathcal{P}_{\delta}(\delta \ln \rho;T)$ with differing $T$, 
 this is analytically unfeasible. Instead, the true PDF can be well 
 approximated by taking the same algebraic form of the PDF Eq.~\eqref{eq:Phils}, 
 with  mean and variance calculated from the true random process    \citep{LPpaper,Hopkins:2015cu}.
 This gives 
 \begin{gather}
 T_{l} = \frac{\int_{\ln l}^{\ln L} T(l') [1+T(l')] d\ln l'}{\int_{\ln l}^{\ln L}[1+T(l')] d\ln l'},\nonumber \\
 \lambda_{l} = \frac{\left(\int_{\ln l}^{\ln L} [1+T(l')] d\ln l'\right)^{2}}{\int_{\ln l}^{\ln L}T(l')(1+T(l')) d\ln l'},\label{eq:T integrals}
\end{gather}
where $T_{l}$ and $\lambda_{l}$ are the ``averaged'' values to use in the PDF Eq.~\eqref{eq:model},
while $T(l) = \kappa(1-\mathcal{M}_{l}^{-2}) = \kappa(1-\mathcal{M}^{-2}L/l)$ is the ``local''
value of $T$ taken from the shock model. Although the integrals in Eq.~\eqref{eq:T integrals} are straightforward
analytically, the added complexity makes these forms inconvenient except for plotting.
In any case, at high $\mathcal{M}$
the differences  compared to the $T=\mathrm{constant}$ assumption of Eq.~\eqref{eq:model}  are  modest, since 
$\mathcal{M}_{l}^{-2}$ falls off  steeply away for  $l>l_{\mathrm{sonic}}$.

\subsection{Choices made and other possibilities}\label{sub:model choices}
We have endeavored through the previous sections to formulate a model with as few parameters as possible,  based on simple physical considerations. 
Indeed,  Eq.~\eqref{eq:model} involves just two physically motivated free parameters, $\kappa$ and $\xi$, to describe 
the variation in a function, $\mathcal{P}_{l}(\ln \rho)$, across all  scales in the system, for 
a wide range of turbulent Mach numbers.
We shall see below (Sec.~\ref{sec:Numerics}) that the model works relatively well in comparison to numerical simulations, both for measures that consider the variation in global parameters with physical parameters (e.g., the variance--Mach-number relation) and for measures in individual simulations (e.g., power spectra and the density PDF). 

 However,  
we feel it useful to reiterate the choices that have been made
throughout the derivation, since some of these can be evaluated directly from simulation (or perhaps observational data). In this 
way, one might imagine calibrating certain aspects of the model,
for example, to improve the accuracy of star-formation models. 
Given the success of the general shape of the PDF \citepalias{Hopkins:2013hn}, here
we consider various aspects of the model that might be modified (and the possible utility in doing so), while
retaining the compound-log-Poisson structure.
\paragraph*{The single-step PDF}
{The choice of an exponential PDF for the size of a single jump $\mathcal{P}_{\delta}$ (Fig.~\ref{fig:stepPDF})
was in part arbitrary, for the sake of convenience. For example, a similar $\mathcal{P}_{\lambda}$ is  obtained with a $\delta$ function
(this is used in most subsonic intermittency models; \citealt{She:1994fn,Boldyrev:2002uh,Hopkins:2015cu,Mallet:2016}), 
or with a shifted Gamma distribution. There are, however, some important properties, which, if not satisfied
would cause
$\mathcal{P}_{l}(\ln \rho)$ to look quite different. In particular, the presence of an 
absolute maximum for $\delta \ln \rho$ is important, since without this $\mathcal{P}_{l}(\ln \rho)$ 
extends to infinitely high densities. Further, if this maximum is not equal to $\epsilon$, the value of 
$\delta \ln \rho $ when there is no event, one has a similar problem, since as $\Gamma \rightarrow 1$ there is a nonzero 
probability  of an arbitrarily large number of events $n$. 
 } 
  \paragraph*{Number of structures encountered}
  {The proportionality between the probability of an event $a$ and
 the jump size in scale $\delta r = \ln\Gamma$ is an important parameter
 that controls the level of intermittency for a given variance (i.e., the relation between $S_{\ln \rho}$ and $T$).
  In our model this is set by relating $a$ to $\epsilon$ using $\langle \rho \rangle=1$, then setting
 $\epsilon$ to the maximum possible density possible from a volume change in  $\xi$ dimensions
  $\epsilon = \xi \delta r$. While the proportionality between $\epsilon$ and $\delta r$ simply relies on having a density field that is approximately scale invariant, the value for $\xi$ is less well constrained. It  is reasonable to expect $\xi <3$, although its exact value depends on properties of the density field such as the density in ``unshocked'' regions. 
  We have found empirically that $\xi\approx 1.3\rightarrow 1.5$  
gives a more accurate match to data (the exact value is hard to constrain given the significant scatter seen in numerical results; see Fig.~\ref{fig:dens variance}),  which is presumably related
to the maximum density being closer to $L/l$ as opposed to $(L/l)^{3}$ since 
compression happens primarily along one dimension (in other words, the maximum possible density is when all of the mass is contained within a single 2-D shock).
  }
 \paragraph*{Variation of $T$ with scale}{
 The variation in $T$ with scale, and/or with Mach number, is difficult 
 to constrain precisely, primarily because the system is \emph{not} scale invariant at moderate Mach number. Specifically, although the sonic scale $l_{\mathrm{sonic}}$ is the {only} physically important scale above
 the viscous scale (which is ideally well into the subsonic regime), and 
 so shock widths should  eventually scale with $l_{\mathrm{sonic}}$, it is hard to know how far above $l_{\mathrm{sonic}}$ it is necessary to go before this occurs. This implies that the shock width could  deviate from $\sim \kappa \mathcal{M}^{-2}$; in other words, $\kappa$ could depend on $\mathcal{M}$  at moderate $\mathcal{M}$. Nonetheless, the general form of $T$---an increase at low $\mathcal{M}$, followed by a flattening at high $\mathcal{M}$---is quite robust. 
 Based on fits to numerical simulations and power spectra (see Secs.~\ref{sub:numerical PDFs} and \ref{subsec:spectra}), we have found that
 the form \eqref{eq:T}  slightly underpredicts the increase in $T$ with $\mathcal{M}$; i.e., it should asymptote more slowly to $T=\mathrm{const.}$. Physically, 
 this implies that the fraction of the mass contained in individual shocks should increase with $\mathcal{M}$ at higher values of $\mathcal{M}$ than suggested by $r_{\mathrm{shock}} = \kappa l_{\mathrm{sonic}}$.

Another possible uncertainly stems from our assumption that the velocity scales as $v_{l}/c_{s}\sim \mathcal{M}(l/L)^{\zeta}$ with $\zeta=1/2$, and that shocks are 2-D structures. If these parameters differ from these fiducial values (as may be the 
 case; see, for example \citealt{Federrath:2008ey,Federrath:2013gu}), this will also change the variation of  $T$ with $\mathcal{M}$.
The sign of the change is such that an increase in shock dimension,\footnote{This can take on noninteger values if the shocks have a fractal structure on scales $l\gg l_{\mathrm{sonic}}$.} or an increase in $\zeta$, causes
 $T$ to increase  with $\mathcal{M}$ (i.e., $T$ changes from being constant at $\mathcal{M}\gg1$ to being a slowly
 increasing function of $\mathcal{M}$, although it must eventually flatten out).}

\vspace{0.3cm}
Finally, it worth emphasizing 
why it can be hard to describe the full density statistics with two parameters. The primary difficulty is that, unlike models that 
assume log-normal statistics, the density variance and its variation with scale 
are now entangled with the distribution's intermittency. For example, the prediction 
for the density-variance--Mach number relation depends strongly on $T$ (or $\kappa$) and its variation with $\mathcal{M}$. Thus, any attempt to increase 
the variance by increasing $\kappa$ creates unphysically intermittent density
distributions. One can partially compensate for this by modifying $\xi$, higher
values of which will increase the variance without much changing the intermittency, but this only works to some degree. Further, it is then necessary to match any measurements to different simulations across a variety of Mach numbers. In this regard, similar models for subsonic turbulence (e.g., \citealt{She:1994fn}) are more easily constrained: there are more known parameters (e.g., the power spectrum), and any model is required to fit only one simulation (there is an inertial range, so there is no requirement for a model that remains accurate across all Mach numbers).
In the tests of the following section, we compare directly to data from numerical simulations. The most stringent of these tests is the explicit calculation of $\mathcal{P}_l(\ln \rho)$ in Sec.~\ref{subsec: PDF as a function of l}, which probes  the variation in both density variance and intermittency with scale.

% Finally, it is worth mentioning our neglect of subsonic contributions; i.e., the 
% approximation $T\approx 0$ for $\mathcal{M}<1$. 
% Including such contributions will add variance to $\mathcal{P}_{\lambda}$
% without contributing much to the intermittency (since the jump size, $T$, will get smaller).
% The neglect of this is primarily for simplicity and philosophical reasons---we prefer
% to focus of the high-$\mathcal{M}$ limit, since the turbulence cannot be scale-invariant across the transition from $l>l_{\mathrm{sonic}}$ to $l<l_{\mathrm{sonic}}$. Nonetheless, we shall see in Sec.~\ref{sub:numerical PDFs}
% that the contribution to the density variance from subsonic scales is minor in comparison to the supersonic 
% contribution as expected. 

\section{Predictions and numerical comparisons}\label{sec:Numerics}

\begin{figure}
\begin{center}
\includegraphics[width=1.0\columnwidth]{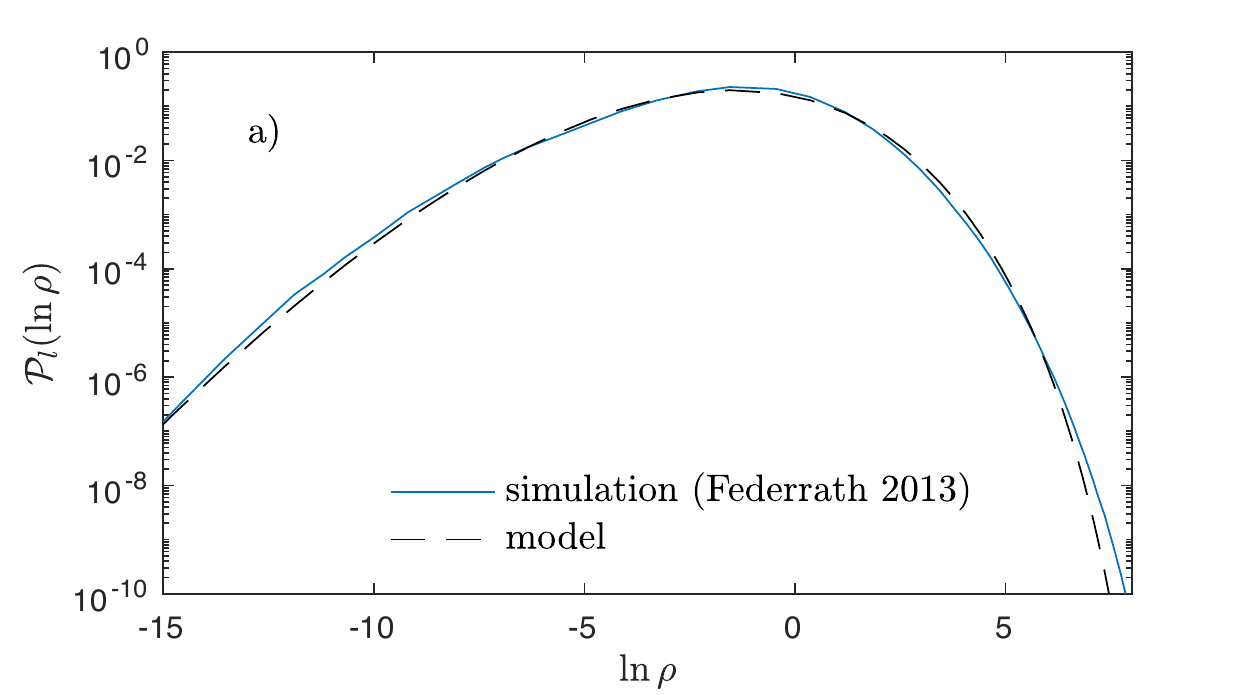}
\includegraphics[width=1.0\columnwidth]{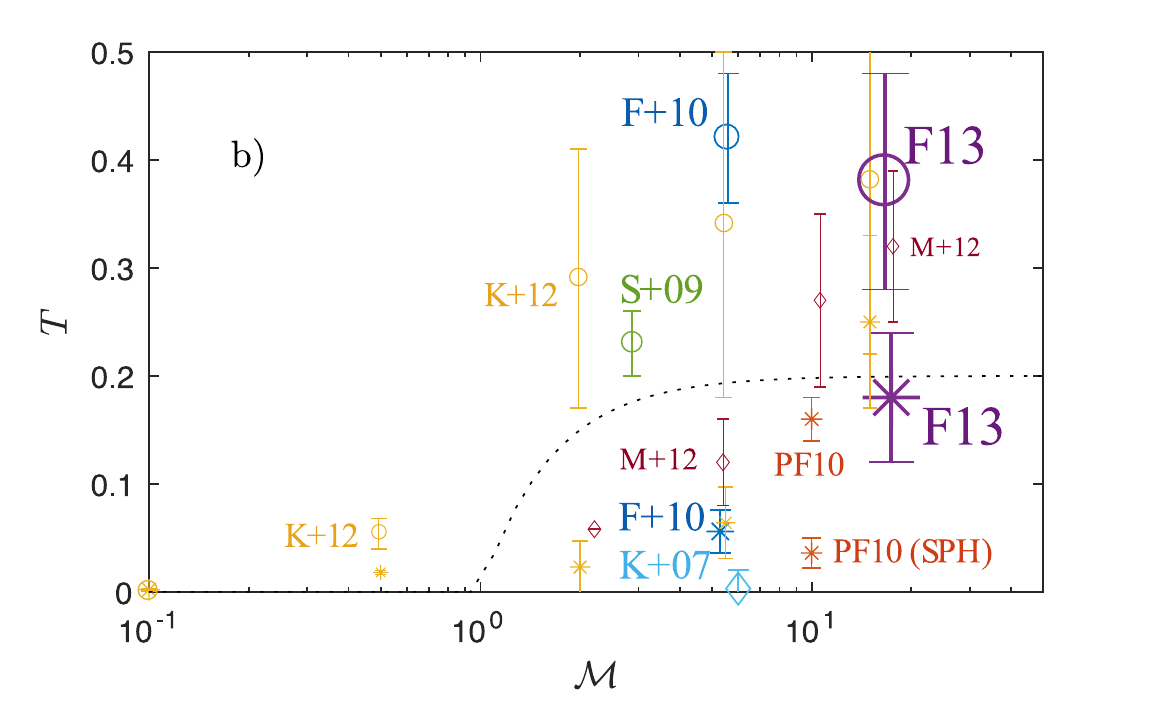}
\caption{(a) Comparison of the model (dashed black line) to the volume-weighted density PDF from the highest resolution simulation currently available (blue line), that of \citet{Federrath:2013gu} at $4096^{3}$.  The simulation uses solenoidal 
forcing and has a Mach number of $17.4 \pm 1.1$, and the fit shown has parameters $\kappa=0.24$, $\xi=1.5$ [this corresponds to $T=0.2$, $S=4.1$ using Eq.~\eqref{eq:T integrals}]. \citet{Federrath:2013gu}
also presents a compressibly forced simulation at similar Mach number, which causes a density
PDF of much higher intermittency $T\approx 0.4$ and is also fit very well by model (see \citealt{Federrath:2013gu} Fig.~4). (b) Measurements of $T$
taken from a variety of simulations as a function of Mach number. Circle markers indicate simulations
with compressive forcing, asterisks indicate solenoidal forcing, and diamond markers indicate forcing
that involves some mix of compressive and solenoidal modes.   Marker sizes are scaled by simulation resolution (a simulation at $N^{3}$ has a marker size $\propto \sqrt{N}$) so as to emphasize the most significant data points.
In addition to the $4096^{3}$ simulations of \citet{Federrath:2013gu} (purple, labelled F13), this data 
is taken from \citetalias{Hopkins:2013hn} based on \citet{Federrath:2010ef} (blue, labelled F+10, $1024^{3}$), \citet{Price:2010ds} (red PF10, $512^{3}$; SPH refers to their smoothed-particle hydrodynamics simulation), \citet{Konstandin:2012ei} (yellow K+12, $512^{3}$), \citet{Schmidt:2009dr} (green S+09, $768^{3}$), \citet{Kritsuk:2007gn} (light blue K+07, $1024^{3}$), \citet{Molina:2012iv} (maroon M+12, $256^{3}$). \citetalias{Hopkins:2013hn} lists the relevant parameters for each simulation in Table~1 (see also  \citetalias{Hopkins:2013hn} Fig.~3).  The dotted line shows the function $T=\kappa(1-\mathcal{M}^{-2})$ for $\kappa=0.2$, which 
 was proposed in Sec.~\ref{sec:model} based on the physical size and density contrast of shocks. 
Note that compressible forcing leads to higher intermittencies ($T$). With the large scatter in the data, it is unclear whether the simple model for $T$ proposed in Sec.~\ref{subsec:T discussion} is correct, but there does not appear to be a strong further increase in $T$ with $\mathcal{M}$ for  $\mathcal{M}\gtrsim 7$ (see Sec.~\ref{sub:model choices}).}
\label{fig:PDF and T}
\end{center}
\end{figure}

In this section, we outline the main predictions of the model, comparing these
to   results from previous works and several numerical simulations. In addition to the basic 
functional form of the PDF, which is well-known to accurately match simulations \citep{Hopkins:2013hn,Federrath:2013gu,Konstandin:2015kv}, the model's scale dependence implies we can predict measures of 
the statistical variation with scale, such as power spectra and structure functions. Each of these is agrees within uncertainty to results from numerical simulations.
In addition, in App.~\ref{app:shock sizes} we test the shock-width hypothesis $r_{\mathrm{shock}}\sim \kappa l_{\mathrm{sonic}} $, to check basic consistency with the model for $T$ laid out in Sec.~\ref{subsec:T discussion}.

\subsection{Density PDF: basic form}\label{sub:numerical PDFs}

The shape of the density PDF is shown for illustration purposes in Fig.~\ref{fig:PDF and T}(a), which 
compares the form \eqref{eq:Phils} to the numerically measured density PDF from the highest-resolution isothermal supersonic turbulence simulation yet run, from \citet{Federrath:2013gu}. As also shown in  \cite{Hopkins:2013hn} for a range of other simulations, the quality of the agreement with the numerical PDF is impressive, with close-to-perfect agreement
seen far into the tails of the distribution where the deviation from log normality is very significant.

The variation of the intermittency parameter $T$ with Mach number is the first prediction of the model beyond 
\cite{Hopkins:2013hn}. In particular, we predict that $T$ should be approximately constant with $\mathcal{M}$ for 
$\mathcal{M}\gg 1$, but may depend on the degree of compressibility (i.e., the ratio of solenoidal to 
compressive motions) of the turbulence, because this will change the shock density contrast $b$ and/or shock width (compared to $l_{\mathrm{sonic}}$). As shown in Fig.~\ref{fig:PDF and T}(b), these behaviors are indeed observed in simulation data (insofar as the very 
large uncertainties permit). In particular, $T$ is very small for $\mathcal{M}<1$, rises rapidly to $\mathcal{M}\sim 3\rightarrow 5$, then appears to stay constant between $0.15$ and $0.4$  above this (possibly with some slow increase with $\mathcal{M}$, more simulations are needed to address this more accurately). In addition, 
the $4096^{3}$ simulations of \citet{Federrath:2013gu} show a large difference in $T$ between 
solenoidally and compressively forced simulations ($T\approx 0.2$ and $T\approx 0.4$ respectively) as expected from the arguments in Sec.~\ref{subsec:T discussion}. 

It is worth noting that the intermittency ($T$) measured from simulations can vary significantly with numerical resolution and the numerical method. For example, Fig.~5 of \citet{Federrath:2013gu} illustrates how under-resolved simulations can overestimate $T$, while the \citealt{Price:2010ds} SPH simulations and \citet{Kritsuk:2007gn} adaptive-mesh-refinement simulations produce very low intermittencies [see Fig.~\ref{fig:PDF and T}(b), points PF10 (SPF) and K+07].

 \begin{figure}
\begin{center}
\includegraphics[width=1\columnwidth]{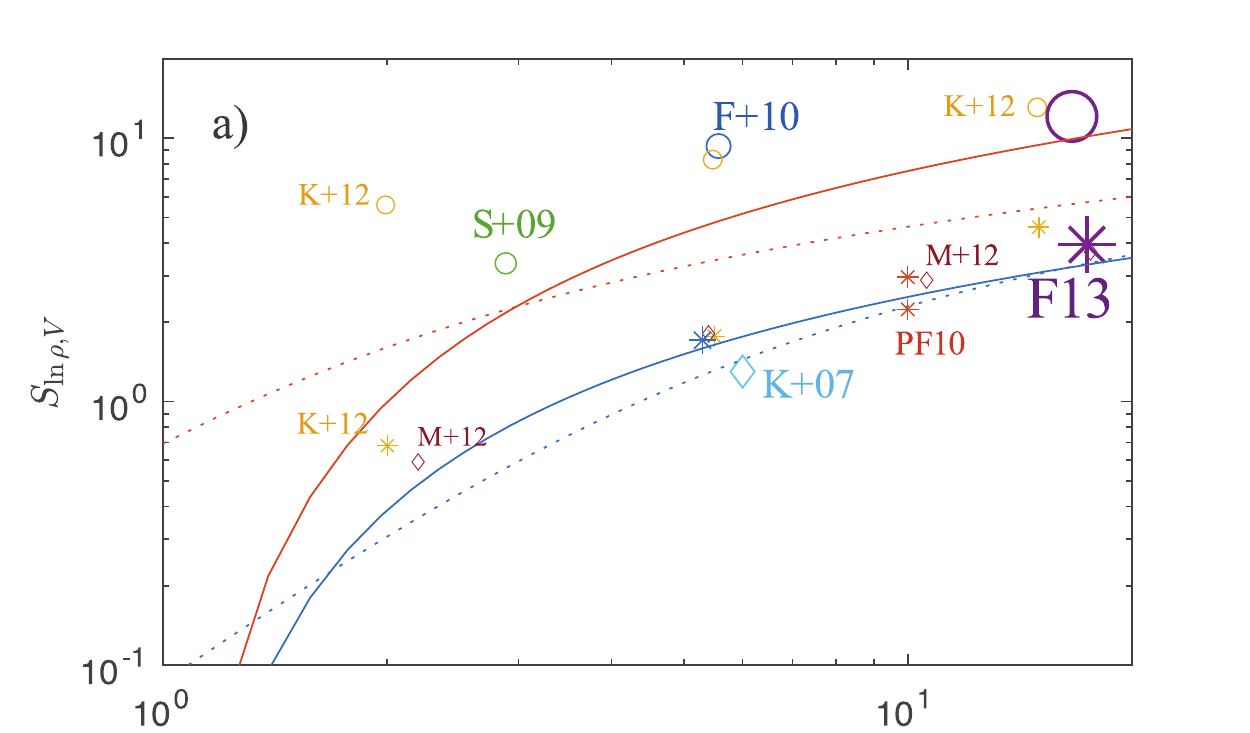}
\includegraphics[width=1\columnwidth]{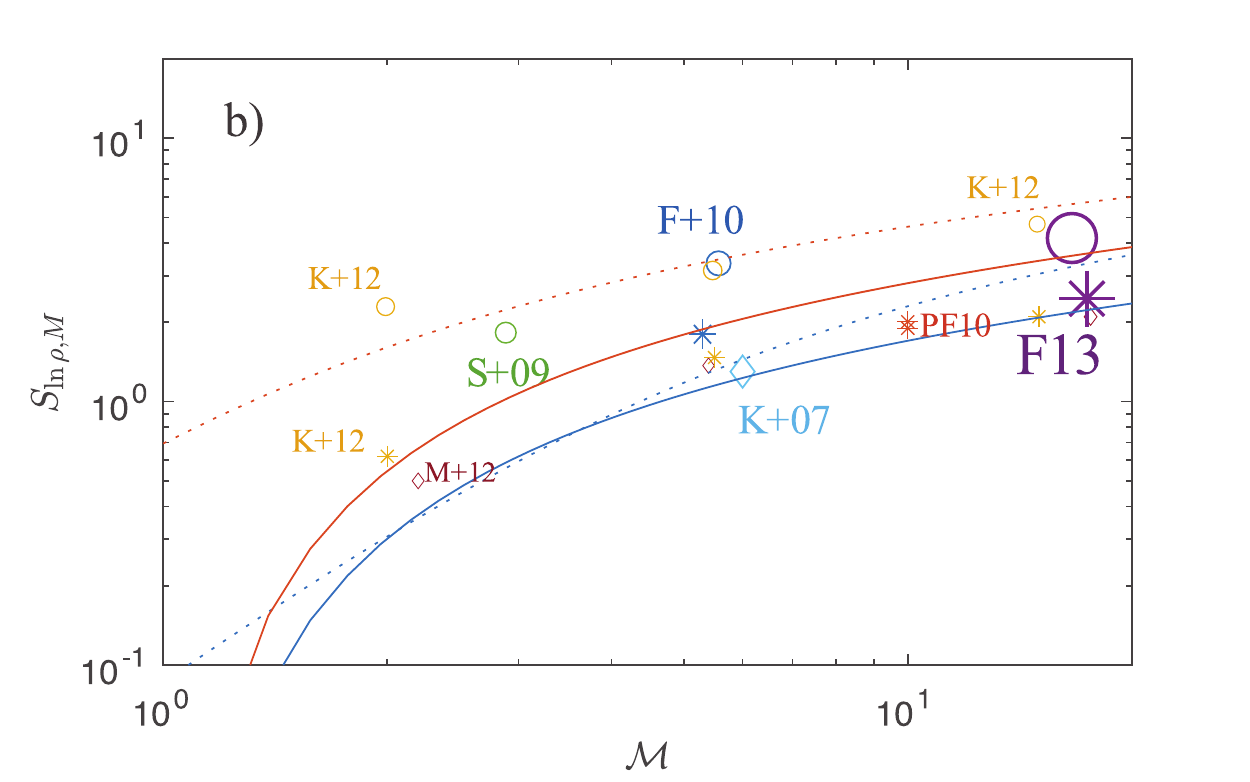}
\caption{Density variance-Mach number relation (for $\mathcal{M}>1$) as predicted by the model (solid lines) and as measured from simulations in 
previous literature (markers). Panel (a) shows the volume-weighted variance $S_{\ln \rho, V}$, while panel (b) shows the mass-weighted variance $S_{\ln \rho, M}$. We show  model predictions for $\xi=1.5$ and $\kappa=0.2$ (blue, lower curve) or $\kappa = 0.6$ (red, upper curve) to indicate a range
of values that might apply to solenoidally and compressively forced turbulence.
% Explain kappa value somewhere =>T a bit less ~0.4 for 
 The dotted curves  show the standard result $S_{\ln \rho} \approx \ln(1+b^{2}\mathcal{M}^{2})$ \citep{Padoan:1997bh} for $b=0.3$ (blue) and $b=1$ (red), respectively for solenoidal and compressive forcing  \citep{Federrath:2010ef}. Note that for the standard result, the volume-weighted and mass-weighted 
variances are the same since the PDF is assumed to be log-normal.
The simulation results are as listed in Fig.~\ref{fig:PDF and T}(b) (see \citetalias{Hopkins:2013hn}), with the marker size again scaled by simulation resolution.
Although there is a large amount of scatter in the simulation results, the proposed model fits the data at least as well as the standard fit and captures the difference between the volume- and mass-weighted distributions.
}
\label{fig:dens variance}
\end{center}
\end{figure}

\subsection{Density variance-Mach number relation}

The density variance-Mach number relation is
the simplest prediction of the model that 
depends on the variation of the PDF with scale (since higher Mach numbers have a larger 
range of scales  $l \gtrsim l_{\mathrm{sonic}}$). Phenomenological arguments and simulations in   previous literature suggest that this relation is  \begin{equation}
S_{\ln \rho}\approx \ln (1+b^{2}\mathcal{M}^{2}),\end{equation}
 with $b\approx 0.3$  solenoidally forced turbulence, or $b\approx 1$ for compressively forced turbulence  (see, for example, \citealt{Padoan:1997bh,Passot:1998cr,Lemaster:2008kb,Price:2011ha,Molina:2012iv}).
This ``standard'' result does not include a distinction between the volume-weighted  variance $S_{\ln \rho,V}$ and the 
mass-weighted variance $S_{\ln \rho,M}$, since these are identical when the density is distributed log-normally. 

From Eq.~\eqref{eq:model} and $S_{\ln \rho,V}=2T^{2}\lambda$, it is clear that since $\lambda \propto \ln (L/l)$,
the model predicts $S_{\ln \rho }\propto \ln (\mathcal{M}^{2})$ for $\mathcal{M}\gg 1$ (i.e., once $T\sim \mathrm{const.}$). 
This is a generic consequence of
the constant accumulation of variance as a function of scale in log space, so this  
form is shared between any compound-log-Poisson model with scale-independent parameters. 
More precisely, the ``exact'' result, using $T$ from Eq.~\eqref{eq:T} and Eq.~\eqref{eq:T integrals}, is 
\begin{equation}
S_{\ln \rho,V} = 2\xi \kappa (\kappa +1)\ln (\mathcal{M}^{2}) + \kappa \xi (1-\mathcal{M}^{-2})[2+\kappa(3-\mathcal{M}^{-2})]\label{eq:SV full},
\end{equation}
which becomes $S_{\ln \rho,V} \approx 2\xi \kappa(1+\kappa) \ln (\mathcal{M}^{2})$ for $\mathcal{M}\gg 1$. 
While this form is not identical to $S_{\ln \rho}\approx \ln (1+b^{2}\mathcal{M}^{2})$, it can be 
very similar for $\mathcal{M}\gtrsim 5$ depending on the constant of proportionality. The question
then becomes: For a reasonable  intermittency (i.e., a value of $T$ that matches that measured from the PDF), 
does the model prediction $S_{\ln\rho,V }\approx 2\xi\kappa (1+\kappa)\ln \mathcal{M}^{2}$ [or Eq.~\eqref{eq:SV full}] also 
match the measured $S_{\ln \rho,V}$?

% EXPLAIN SOMETHING ABOUT WHY KAPPA ~0.5

In Fig.~\ref{fig:dens variance}, we compare the model prediction 
with the standard result $S_{\ln \rho}\approx \ln (1+b^{2}\mathcal{M}^{2})$ and various previous simulation results for $\mathcal{M}>1$. This is done for both the volume-weighted variance [Fig.~\ref{fig:dens variance}(a)] and the mass-weighted variance [Fig.~\ref{fig:dens variance}(b)]. We plot $S_{\ln\rho}$ in each case for the model parameters $\kappa = 0.2$ and $\kappa = 0.5$, which are chosen to illustrate a range of intermittencies seen in solenoidally and compressibly forced simulations.\footnote{The values of $T_{l}$ from Eq.~\eqref{eq:T integrals} are somewhat lower than $\kappa$. For example, $\kappa=0.5$ gives an intermittency parameter $T\approx 0.4$ for the compressible simulation of  \citet{Federrath:2013gu}, which 
is close to the measured value.}
Given the significant scatter, the agreement of the model prediction is decent for these reasonable values of $\kappa$, and  it seems fair to say that the model relation is of (at least) a similar quality  to the standard result. 
Both results appear to somewhat underestimate the compressible variance, although more simulations at higher resolution
are needed to better assess the trends at higher Mach number.
We also see better agreement between the predicted and measured mass-weighted 
variances $S_{\ln \rho,M} = (1+T)^{-3}S_{\ln \rho,V}$, compared to the standard result. This is unsurprising since the effects of intermittency generically
act to reduce $S_{\ln \rho,M}$ compared to $S_{\ln \rho,V}$. 
 It is also worth reiterating that the model neglects the subsonic contribution to the variance, which is significant for simulations with   $\mathcal{M}$ approaching $1$, and is the cause of the variance under-prediction at low $\mathcal{M}$.

% \begin{figure}
% \begin{center}
% \includegraphics[width=1.0\columnwidth]{Figures/PDFsScaleM10}
% \includegraphics[width=1.0\columnwidth]{Figures/PDFsScaleM30}
% \includegraphics[width=1.0\columnwidth]{Figures/PDFsScaleM5}
% \caption{\textcolor{red}{REMOVE THIS }Density PDF $\mathcal{P}_{l}(\ln \rho)$ for (a) $\mathcal{M}\approx 12$, (b) $\mathcal{M}=40$, and (c) $\mathcal{M}\approx 7 $. Solid curves in each plot show the measured PDFs on scales $l = L/512$ (black, widest curve; 
% this should be considered the PDF on the smallest scales), $l=L/32$ (blue, middle curve), and $l=L/4$ (red, narrowest curve), using the fiducial maximum density scaling $\epsilon = 3\delta r$ ($\xi=3$). The dashed curves of matching colors show the model predictions for these parameters at $\kappa = 0.11$ (a-b) and $\kappa =0.08$ (c).
% Although the predicted intermittency is somewhat too low (i.e., the predicted distributions
% are not sufficiently skewed to low densities), it is encouraging that very 
% similar values for $\kappa$ give decent results across a wide range in Mach number. Note that
% the $\mathcal{M}\approx 7$ simulation is more affected by the subsonic contribution, 
% which would cause a larger variance on smallest scales $l=L/512$, and  explains
% the mismatch in the high-density tail [see Fig.~\ref{fig:PDF scales 1.3}(c)].}
% \label{fig:PDF scales}
% \end{center}
% \end{figure}

\subsection{Density PDF as a function of scale}\label{subsec: PDF as a function of l}
A more stringent test of model predictions is to compute $\mathcal{P}_{l}(\ln \rho)$ 
directly from simulation, viz., bin  the density into volumes of size $l^3$ then compute the density PDF. 
%In fact, in addition to being an important property of the turbulence for star-formation/fragmentation
% applications,
%one might argue that the density PDF averaged over the sonic scale $l=l_{\mathrm{sonic}}$ is
%a more physically relevant quantity than the grid-scale PDF for basic turbulence studies, being free 
%of subsonic contributions that stem from different physical processes. 
Unfortunately, so far as we are aware there is no measurement of this in previous literature, despite its  physical, as well as theoretical, relevance.
We have thus run a variety of  isothermal turbulence simulations to make such measurements directly.
These simulations use the GIZMO code \citep{Hopkins:2015bk,Hopkins:2016ik} with the Meshless-Finite-Mass (MFM) method and $256^{3}$ elements. 
Although this resolution may be relatively low by modern standards, the Lagrangian nature of 
the MFM method more accurately captures the high density shock regions by naturally 
having higher resolution in such regions \citep{Price:2010ds}, and the MFM method has proven 
very accurate in a wide variety of test problems \citep{Hopkins:2015bk,Hopkins:2016ik}.
That said, given the significant dependence of intermittency properties on resolution and numerical method
\citep{Price:2010ds,Federrath:2013gu}, it will be important to verify the scaling of these results with resolution.
The simulations are forced by   a Ornstein-Uhlenbeck process 
with an equal mix of solenoidal and compressive large-scale modes, as described in \citep{bauer:2011.sph.vs.arepo.shocks}.
Different forcing strengths are used to drive turbulence across a range of  Mach numbers.
PDFs are calculated by depositing the density field onto a 
$512^{3}$ uniform grid using a Gaussian kernel for each Lagrangian mesh element (of width $\sigma =\sqrt{3/40} h$, where $h$ is the cell smoothing length; \citealt{2012MNRAS.425.1068D,Hopkins:2015bk}), then averaging over successively larger  
volumes to form the PDF as a function of scale. This grid-based
method agrees with the volume-weighted PDF calculated directly from the Lagrangian mesh
for the finest  $N=512$ grid.\footnote{There is some discrepancy at the lowest densities. This is expected 
because the density field deposited using the Gaussian kernel contains 
regions of  lower density (the regions in between mesh elements) than that of the lowest density mesh elements.
 To find the true density in such regions
one should use the ``gather'' method for constructing gridded data, as this is
actually used in the simulation \citep{Hopkins:2015bk}; 
however the Gaussian kernel method we use probably provides a 
better representation of the true field  than the PDF from Lagrangian data, which 
is effectively under-sampling the lowest density regions.
  }

   \begin{figure}
\begin{center}
\includegraphics[width=1.0\columnwidth]{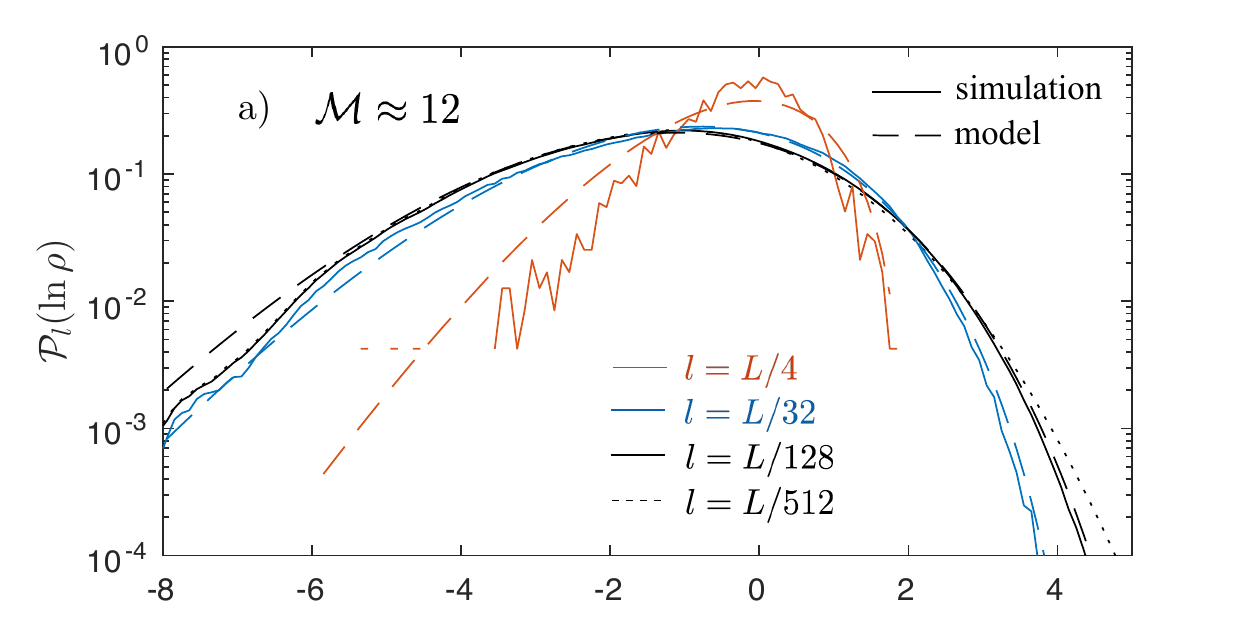}
\includegraphics[width=1.0\columnwidth]{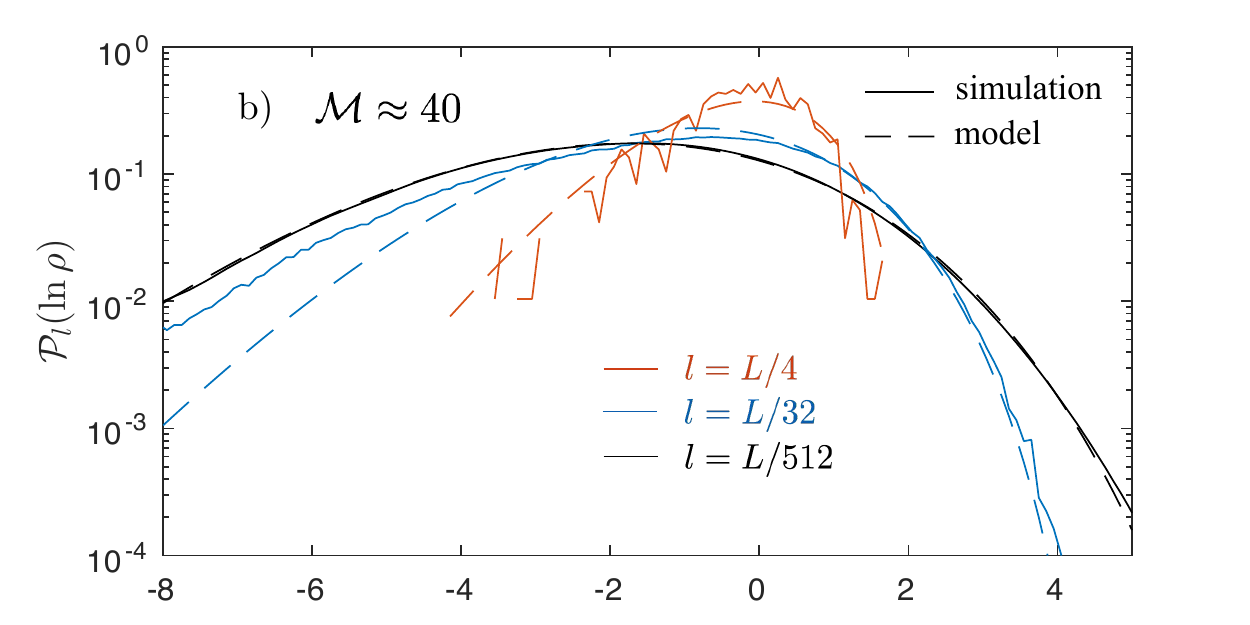}
\includegraphics[width=1.0\columnwidth]{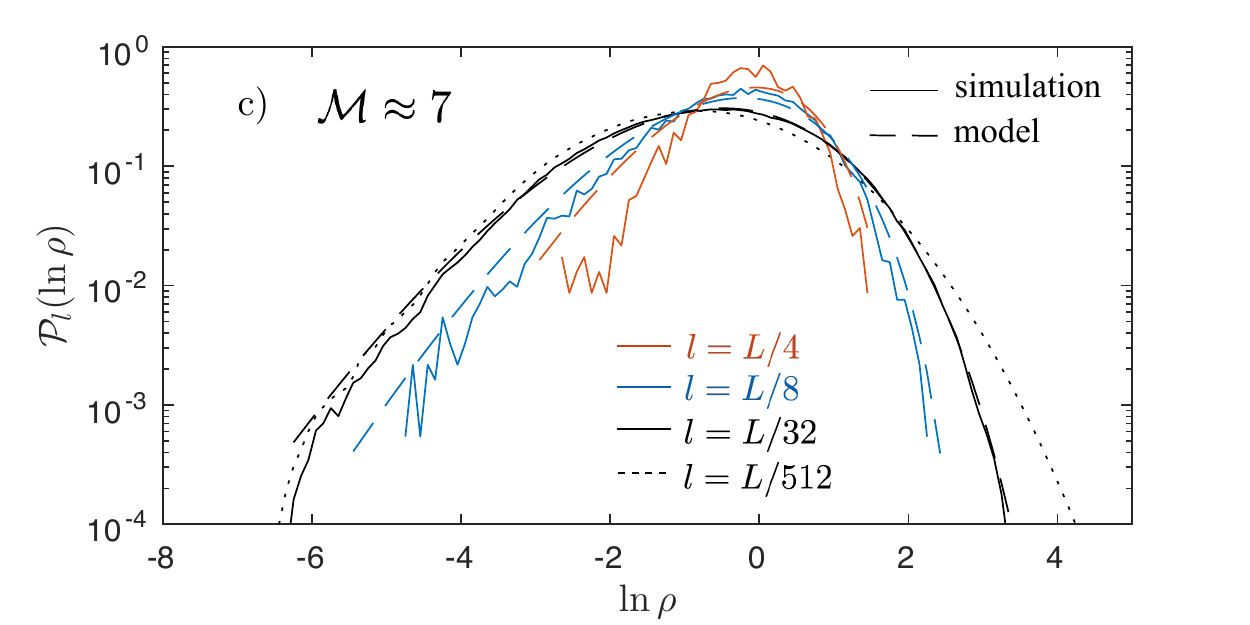}
\caption{Density PDF $\mathcal{P}_{l}(\ln \rho)$ for (a) $\mathcal{M}\approx 12$, (b) $\mathcal{M}\approx 40$, and (c) $\mathcal{M}\approx 7 $ turbulence. In each panel, solid curves show the measured PDFs of the density averaged over a variety of scales, while dashed curves show model predictions for the same parameters. In each case, we only compare the model with simulations on scales $l>l_{\mathrm{sonic}}\approx \mathcal{M}^{-2}$; specifically for (a) $\mathcal{M}\approx 12$ with $l_{\mathrm{sonic}}\approx L/150$, we take $l=L/128$ (black, widest curve), $l=L/32$ (blue, middle curve), and $l=L/4$ (red, narrowest curve); for (b) $\mathcal{M}\approx 40$ with $l_{\mathrm{sonic}}\approx L/1600$, we take $l=L/512$ (black), $l=L/32$ (blue), and $l=L/4$ (red); and for (c) $\mathcal{M}\approx 7$ with $l_{\mathrm{sonic}}\approx L/50$, we take $l=L/32$ (black), $l=L/8$ (blue), and $l=L/4$ (red). In this way
the subsonic contributions to the variance, which are not included in the model, are explicitly removed from the simulation results. We take $\kappa = 0.24$ (a-b) and $\kappa = 0.18$ (c), with $\xi =1.3$ in all cases,  illustrating its 
success across a range of $\mathcal{M}$ with little change to $\kappa$ (see text for discussion).
In each panel, we also show the PDF on the smallest scales measured, $l = L/512$, and the difference between this and the solid 
black curve explicitly illustrates the contributions from subsonic scales in  (a) and (c).
Although there seems to  be a slight under-prediction of the intermittency on the largest scales $l=L/4$,
the numerical PDF is likely also affected by discretization and insufficient statistics, which would tend 
to raise the low-density tail.}
\label{fig:PDF scales 1.3}
\end{center}
\end{figure}
 
 Figure~\ref{fig:PDF scales 1.3} compares  results from simulations at (a) $\mathcal{M}\approx 12$, (b) $\mathcal{M}\approx 40$, and (c) $\mathcal{M}\approx 7 $ with model predictions (shown with dashed lines). 
In each case,  to specify the model, we use the measured $\mathcal{M}$, $\xi=1.3$, and  choose $\kappa$ to match the  variance on the smallest supersonic scales.\footnote{This value of $\xi=1.3$ works slightly better for these simulations than the $\xi=1.5$ plotted in Fig.~\ref{fig:dens variance}. Given the large scatter across different simulations and numerical methods, it is not surprising that model parameters should need to be modified slightly to fit a particular simulation set.} We then compare the model predictions for the intermittency and variance on scales  $l>l_{\mathrm{sonic}}$ with that measured by binning 
the density  on  the same scale (for three values of $l$). More precisely, in the $\mathcal{M}\approx 12$ simulation [Fig.~\ref{fig:PDF scales 1.3}(a)], $l_{\mathrm{sonic}}\approx L/150$, so we compare data on scales $l = L/128$, $l = L/32$, and $l = L/4$;  in the $\mathcal{M}\approx 40$ simulation [Fig.~\ref{fig:PDF scales 1.3}(b)],  $l_{\mathrm{sonic}}\approx L/1600$, so we  compare data on scales $l = L/512$, $l = L/32$, and $l = L/4$; and in the $\mathcal{M}\approx 7$ simulation [Fig.~\ref{fig:PDF scales 1.3}(c)],  $l_{\mathrm{sonic}}\approx L/50$, so we  compare data on scales $l = L/32$, $l = L/8$, and $l = L/4$.  This method is chosen to explicitly remove the subsonic 
scales from the comparison, since these are not included in the model. 

The agreement of the model to simulation is seen to be relatively good. In particular, identical model parameters ($\kappa=0.24$ and the measured $\mathcal{M}$)  give very good fits to  $\mathcal{M}\approx 12$ and $\mathcal{M}\approx 40$ across a wide range of scales in the system. A similar, though slightly lower,  value of $\kappa$ ($\kappa=0.18$) gives a very good fit at lower $\mathcal{M}\approx 7$. We attribute this difference in $\kappa$ to the slight underestimation  of the increase in $T$ with $\mathcal{M}$ in  the model [Eq.~\eqref{eq:T}]. Although the model possibly overpredicts the variance at the largest scale  ($l=L/4$) in each case, it is worth noting that this scale is very  close to the driving (at $\sim L/2$) and may be influenced by this. 
Further,  the statistics at $l=L/4$ in each case are somewhat undersampled (there are only 
$64$ values per density snapshot),  and more values would tend to increase the low density tail.

Finally, it is worth briefly mentioning the contribution of the subsonic scales to the full $512^3$ density PDF (dotted line in each panel of Fig.~\ref{fig:PDF scales 1.3}).
As can be seen from Fig.~\ref{fig:PDF scales 1.3}(a) and (c),  the subsonic  scales 
have the effect of decreasing the intermittency (i.e., making the distribution more log-normal) by contributing to the high-density tail.  This  should be expected, since the subsonic contribution
will involve large numbers of small events (i.e., small $T$). While the effect appears  more 
significant in the $\mathcal{M}\approx 7$ simulation (compared to the larger $\mathcal{M}$ cases), in fact, the absolute increase in the density maximum---i.e., 
the difference between $\rho_{\mathrm{max}}$ with and without the subsonic contributions---
is about the same at  $\mathcal{M}\approx7$ and $\mathcal{M}\approx12$. It appears larger at $\mathcal{M}\approx7$ due to the smaller contribution to the variance from supersonic motions. This justifies our neglect of subsonic scales in the model, which is primarily intended for study of the $\mathcal{M}\gg 1$ limit.  Since the subsonic contribution occurs on the very smallest scales 
of any simulation, and will thus presumably be affected by the numerical method, it is possible that they play a role in the wide 
scatter seen between different simulations in both the density variance-Mach number relation and the intermittency (see Figs.~\ref{fig:dens variance} and \ref{fig:PDF and T}). See \citet{Federrath:2010ef} for further discussion.

\subsection{Spectrum}\label{subsec:spectra}

As shown in App.~\ref{app:PS derivation}, the density power spectrum $\phi_{\rho}(k)$ is related to the variation  in the second order 
statistics (variance and mean) of the PDF with scale. In particular, 
for some variable $s$, the 1-D power spectrum is 
\begin{equation}
\phi_s(k) \sim  \frac{d}{dk} (S_l + \bar{s}_l^2),
\end{equation}
where $S_{l}$ and $\bar{s}_{l}^{2}$ are the variance and mean of $\mathcal{P}_{l}(s)$.
From Eq.~\eqref{eq:Phils}, the volume-weighted variance of $\rho $ can be calculated as \citepalias{Hopkins:2013hn},
\begin{equation}
S_{\rho} = \exp\left(\frac{S_{\ln\rho}}{(2T+1)(T+1)}\right)-1.\end{equation}
Then, using $\langle\rho\rangle =1$, the simplified form for $S,$ $S_{\ln\rho} = 2 \xi T(1+T) \ln(L/l)$ (i.e., 
neglecting the scale variation of $T$),\footnote{The
power spectrum can be derived analytically using the full integrals, Eq.~\eqref{eq:T integrals}; 
however the resulting expressions are very complicated and no longer 
follow a power law at high $k$.}  and approximating $T\approx \kappa$ for $\mathcal{M}\gg1$, one
obtains the scaling
\begin{equation}
S_{\rho,l}+\bar{\rho}_{l}^{2} \sim \left(\frac{l}{L}\right)^{-\frac{2\xi \kappa}{2\kappa+1}}.\end{equation}
The model thus predicts the $\rho$ power spectrum
\begin{equation}
\phi_{\rho}(k) \sim k^{-\nu}\,\quad \nu = \frac{1+2\kappa(1-\xi)}{1+2\kappa},\label{eq:power law rho}
\end{equation}
in the limit $\mathcal{M}\gg 1$ for $1/k \gg l_{\mathrm{sonic}}$. This spectrum 
is somewhat less steep than $k^{-1}$ (it is $k^{-1}$ $\kappa\rightarrow 0$ or $\xi\rightarrow 0$) 
and becomes less steep
with increasing $\kappa$ or $\xi$---for example, the parameters used in Fig.~\ref{fig:PDF scales 1.3}
give $\phi_{\rho}(k)\sim k^{-0.56}$.
The same procedure for the 
power spectrum of $\ln \rho$, using $\langle \ln \rho \rangle = - S_{\ln \rho}(1+T)^{-1}/2$ leads to 
\begin{equation}
\phi_{\ln \rho}(k) \sim \frac{2\xi \kappa}{9}\left(k^{-1}+\frac{3 \ln k}{k} \right)+ \mathcal{O}(\kappa)^{2}, \label{eq:power law ln rho}
\end{equation}
implying the power spectrum of $\ln \rho$ is not expected to be a power law but is close to $\sim\! k^{-1}$.

An important difference compared to previous 
models of the supersonic density power spectrum \citep{Saichev:1996wj,Kim:2005fq,Konstandin:2015kv}
is that we do not predict a density spectrum that approaches 
$k^{0}$ for $\mathcal{M}\gg 1$. Instead, our prediction is that 
the spectrum approaches some power law  between $k^{0}$  and $k^{-1}$ that depends relatively strongly on the intermittency of the density distribution (through $\kappa$). We thus predict a steeper spectrum for solenoidal compared to compressive forcing (since measured values of $T$ are larger for compressive forcing); e.g., using the values $\kappa \approx 0.2$ and $\kappa \approx 0.5$ (with $\xi=1.3$) suggests the spectra $\sim k^{-0.63}$ and $\sim k^{-0.35}$ for solenoidal and compressive forcing respectively.
We also predict that the $\ln \rho$ spectrum should not depend 
on the forcing  so strongly, although it is also not a power law. 
These predictions are valid only for scales well above the sonic scale, since 
model parameters change significantly as $\mathcal{M}_{l}\rightarrow 1$.

With currently available simulation data, these predictions are difficult 
to verify or disprove. While a number of studies have considered density
  power spectra across a range of $\mathcal{M}$ \citep{Kim:2005fq,Kritsuk:2007gn,Kowal:2007gz,Konstandin:2015kv},
 the scaling exponent
  depends significantly on resolution (see \citealt{Konstandin:2015kv},  Fig.~8).
Nonetheless, the predicted density spectrum scaling does not appear 
to  {disagree} with previous results,
although the spectrum is likely somewhat steeper than predicted at modest  $\mathcal{M}$. This is expected because the additional decrease in $T$  with scale at moderate $ \mathcal{M}$ [see  Eq.~\eqref{eq:T}] is not taken into account in Eq.~\eqref{eq:power law rho} and acts to steepen the spectrum. 
Similarly, the simulations presented in Fig.~\ref{fig:PDF scales 1.3}
have density power spectra consistent with predictions at high $\mathcal{M}$ ($\sim k^{-0.6}$ at $\mathcal{M}\approx 40$ and 
$\sim k^{-0.7}$ at $\mathcal{M}\approx 12$; comparable to the prediction $\kappa=0.24$, $\sim k^{-0.56}$) but are a little steeper at  $\mathcal{M}\approx 7$ ($\sim k^{-0.9}$), presumably due to the scale dependence of  $T$.

% INCLUDE SOMETHING HERE ABOUT HOW CHOOSING A STRONGER FORM OF \kappa WOULD DECREASE PS
The results for the spectrum of $\ln \rho $ are even less well known, 
to our knowledge appearing in previous literature only in \citet{Kowal:2007gz} (for magnetohydrodynamic
turbulence) and \citet{Federrath:2010ef}. 
\citet{Federrath:2010ef} reports $\ln\rho$ spectra of $\sim k^{-1.6}$ and $\sim k^{-2.3}$ for solenoidal and compressible turbulence respectively at $\mathcal{M}\approx 5.5$, while  \citet{Kowal:2007gz} report $\sim k^{-1.5}$ for $\mathcal{M}\approx 7$ (or perhaps flatter at the largest scales); however, 
because these simulations each have a relatively modest Mach number, a spectrum steeper than $k^{-1}$ should be expected since they
are  in the regime where $T$ increases with $l/L$ [i.e., before the plateau in Fig.~\ref{fig:PDF and T}(b)],
which means the accumulation of variance with $\ln(l)$ is faster than linear. 
Our simulations give somewhat unclear results, although  also appear to show  steeper spectra than predicted (somewhere
between $k^{-1}$ and $k^{-2}$ for the largest scales).  

Overall, these results may suggest that the $T(\mathcal{M})$ scaling $T\sim \kappa(1- \mathcal{M}_{l}^{-2})$ [Eq.~\eqref{eq:T}]
underestimates the true increase of $T$ with $\mathcal{M}$ somewhat (see Sec.~\ref{sub:model choices}). For the sake of example, if we assumed a different  model with a faster increase in $T$, $\kappa\sim \mathcal{M}_{l}^{2}\sim (l/L) \mathcal{M}^{2}$ at moderate $\mathcal{M}$, this would steepen the $\ln\rho$ power law to $\sim k^{-2}$, and the power law of $\rho$ by a factor $\sim k^{-1}$.
Study of higher resolution simulations at higher $\mathcal{M}$ (for example, the simulations of \citealt{Federrath:2013gu}) is required to better assess our predictions. Nonetheless, the general 
arguments presented here should provide a useful framework for interpreting results.

%%%%%%%%%%%%%%%%%%%%%%%%%%%%%%
\section{Extensions}\label{sec: Extensions}

Given the simple relation between the intermittency parameter
$T$ and the physical properties of shocks, it is possible to straightforwardly extend
the model to situations with more complex physics. Here, we briefly consider turbulence with a non-isothermal polytropic equation of state, as relevant for  various phases 
of the ISM (see, for example, \citealt{Audit:2005df,Gazol:2013bx,Federrath:2015gc} and references
therein). Other extensions---e.g., 
to supersonic magnetohydrodynamic (MHD) turbulence---are also possible using similar ideas.
Note that the discussion and results in this section are intended to be of a qualitative
nature. While the simple extensions we propose do give a 
reasonable match to simulation results, the purpose of the analysis is as much  to illustrate the applicability of the shock model in Sec.~\ref{sec:model}, as to provide useful models for turbulent PDFs.
With this in mind, some of the ideas discussed can likely be applied more rigorously if so desired; for example, 
to derive scalings for the low or high density tails of the PDF .

\begin{figure}
\begin{center}
\includegraphics[width=1.0\columnwidth]{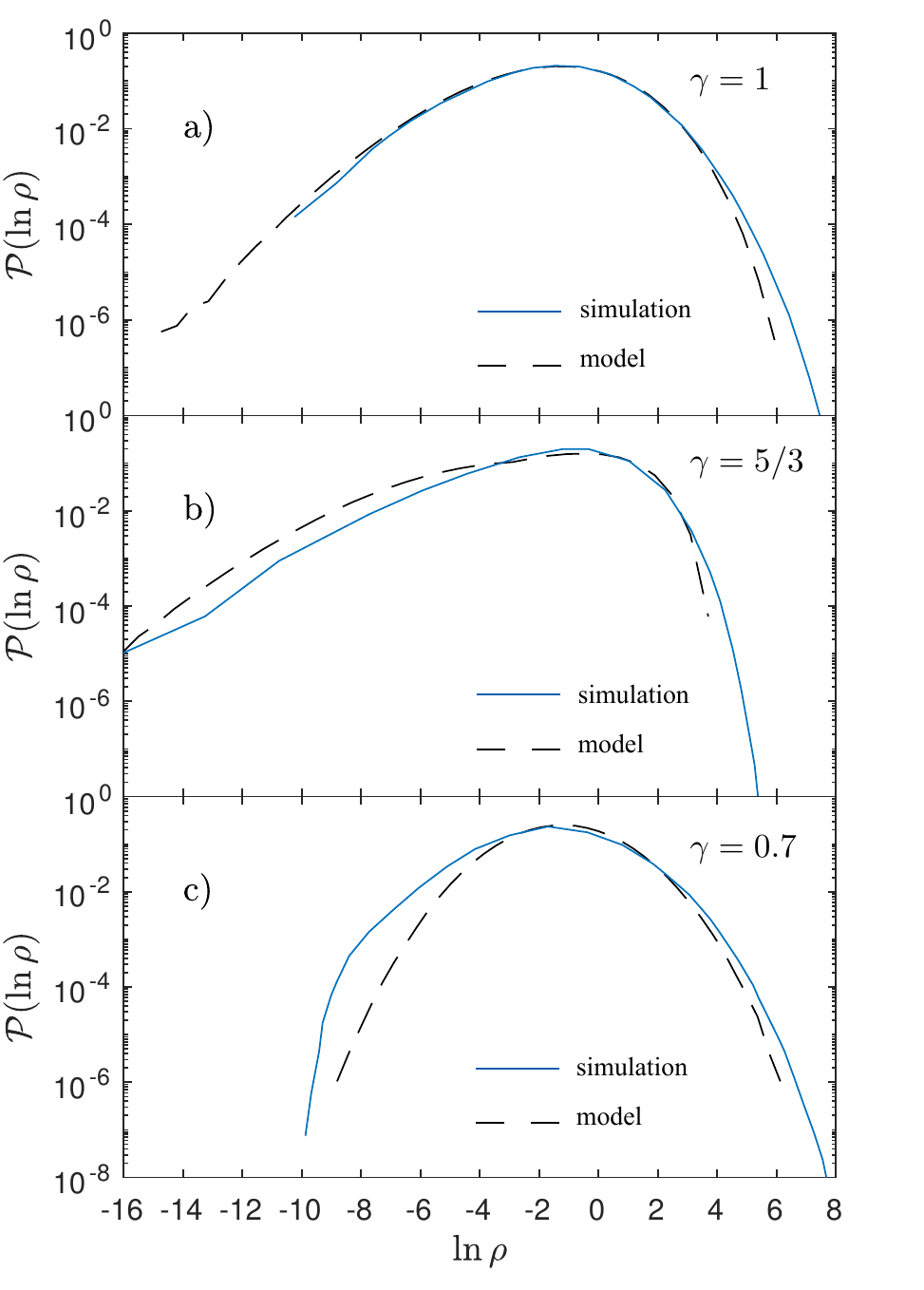}
\caption{
Predicted density PDF for non-isothermal polytropic turbulence (dashed black line) compared to the simulations of \citet{Federrath:2015gc} (solid blue line). The Mach numbers are 
$\mathcal{M} = 11.6$ ($\gamma = 1$), $\mathcal{M} = 13.3$ ($\gamma = 5/3$), and $\mathcal{M} = 8.4$ ($\gamma = 0.7$), and the \citet{Federrath:2015gc} simulations were run at resolutions of $2048^{3}$ ($\gamma=5/3$ and $0.7$)
or $1024^{3}$ ($\gamma=1$). 
For the model, there are \emph{no} free parameters used to fit the non-isothermal PDFs: we use the same value of $\kappa=0.26$ in each case (this was chosen to match the 
$\gamma=1$ distribution) and the physical  value of the Mach number listed above.  
 Although
the fits here are not perfect, the method does a good job at capturing the qualitative 
change in the PDF shape considering there are no free parameters.
In addition, the low density regions where the largest discrepancies are seen have more
significant numerical error bars and resolution dependence  (see \citealt{Federrath:2015gc} Fig.~4), and
there is presumably some contribution to the high densities from the subsonic scales (see, e.g., Fig.~\ref{fig:PDF scales 1.3}(a)).}
\label{fig:non iso}
\end{center}
\end{figure}

The key idea of the method---which is effectively that proposed in  \citet{Passot:1998cr} extended
to  non-lognormal isothermal PDFs---is to assume that the primary effect of 
the non-isothermal equation of state is to modify the sound speed with the 
 density. This in turn modifies the local Mach number and causes the shock density contrast to depend 
on the local value of the density. As shown in \citet{Federrath:2015gc}, this behavior 
is indeed seen as a correlation between $\rho$ and $\mathcal{M}$ in the turbulent joint Mach number-density PDF, as well
as leading to a useful estimate for the modified density variance-Mach
number relation when applied to the shock-jump condition.

More precisely, 
for a polytropic equation of state $p/p_{0} = (\rho/\rho_{0})^{\gamma}$ ($p$ is the pressure), one takes
\begin{equation}
\mathcal{M}\propto c_{s}^{-1}\sim\left(\frac{p}{\rho}\right)^{-1/2} \sim \rho^{(1-\gamma)/2} = \exp\left( \frac{1-\gamma}{2} \ln \rho \right),\end{equation}
which is then used in the shock-density-contrast relation $(\rho_{1}/\rho_{0})\sim b^{2}\mathcal{M}^{2}$.
We thus see that with $\gamma < 1$, the system will have 
higher contrast shocks (compared to isothermal expectations) at high densities and lower 
contrast shocks at low densities, 
while $\gamma > 1$ leads to the opposite behavior.
Within our model, this causes the mean jump size $\langle\delta \ln \rho \rangle -\epsilon = T$ to depend 
on the local value of the density through the replacement of $\mathcal{M}_{l}$ with $\mathcal{M}_{l}\exp[(1-\gamma)/2\, \ln\rho]$, or
\begin{equation}
T\sim \kappa \left(1- \frac{L}{l}\mathcal{M}^{-2} e^{(\gamma -1)\ln\rho}  \right),\label{eq:T noniso}\end{equation}
(with $T=0$ if $\mathcal{M}_{l}^{-2}e^{(\gamma -1)\ln\rho} >1$).\footnote{Note that 
we have neglected a potentially important effect here, which is the change in 
shock contrast and width with $\gamma$ due to the differing sound speeds on either side of the shock. 
The density contrast is derived in the form of a transcendental equation in \citet{Federrath:2015gc}; however, 
the complexity of these expressions, as well as the necessity of deriving the density jump
based on the shock width (which also must scale with the density), leads to complex systems of transcendental 
equations that are difficult to use in the model. The method is thus more similar to \citet{Passot:1998cr}
as opposed to \citet{Federrath:2015gc}. Although 
working this out correctly will certainly change the functional dependence of 
$T$ on $\mathcal{M}$, as well as adding $\gamma$ dependence into the shock jump size ($\kappa$ or 
some similar parameter), the
key differences compared to an isothermal equation of state---in particular the reduction in the size of jumps at low (high) density for $\gamma<1$ ($\gamma>1$)---
are retained in the much simplified version.}
It is clear that this form of $T$ will decrease the low-density tail for $\gamma <1$ due to the fast increase in
$e^{(\gamma-1)\ln \rho}$. In contrast, for $\gamma>1$ the low-density tail will increase in probability,  because the cascade can proceed
further (go to smaller scales) before $T\rightarrow 0$, meaning individual $\delta \ln \rho $ jumps are larger.

Results are illustrated in Fig.~\ref{fig:non iso}, which shows the comparison of this model to the volume-weighted PDF 
data from \citet{Federrath:2015gc}. 
Despite a variety of limitations (see below), we see that the qualitative trends for the PDFs are captured well. 
In particular, at $\gamma>1$ we see a faster fall off at high densities and a long tail at low densities with a  
slope that generally matches the simulation PDF, while at $\gamma < 1$ the lower densities are significantly 
reduced (although the prediction is too severe, cutting off at somewhat higher densities than in the simulation).
Given the possible resolution dependence of the low probability regions (see \citealt{Federrath:2015gc} Fig.~4, second row), the overall agreement is encouraging.

To calculate the illustrated PDFs, we use a
simple Monte-Carlo method
%\footnote{It is possible to derive an approximate analytic prescription by inserting the density dependent PDF into Eq.~\eqref{eq:model}. However, because the true model involves finite sized jumps, this prescription predicts a too sudden fall off in the PDF at  low ($\gamma<1$) or high ($\gamma>1$) densities.It may be possible to correct for this to obtain analytic predictions of the scaling of the tails, but we leave this to future work.}
with the prescription for $T$ taken from Eq.~\eqref{eq:T noniso}, using the same value of $\kappa=0.26$ and $\xi=1.5$ in each case,
with $\mathcal{M}$ as quoted in \citet{Federrath:2015gc}. 
We thus have no free parameters to aid in the  fitting for the nonisothermal PDFs in Fig.~\ref{fig:non iso}, 
and the fit could potentially be significantly improved by optimizing over $\kappa$.
(Note that, based on physical arguments, $\kappa$  \emph{should} be modified somewhat with $\gamma$, becoming smaller
with increasing $\gamma$ due to the
change in shock jump condition calculated in  \citealt{Federrath:2015gc}.)
For consistency with the
simulations, we retain  scales down to $L/l \sim 1024 $ in the Monte-Carlo cascade (estimating twice the grid 
scale as
the minimum resolvable scale). 
In addition, the subsonic scales are included in the 
illustrated PDFs making a direct comparison difficult, and at this Mach number $\mathcal{M}\sim 12$ these will have a minor but observable contribution to the high-density probability (e.g., compare the $L/l=512$ and $L/l=128$ curves in Fig.~\ref{fig:PDF scales 1.3}(a), which explicitly shows the subsonic contribution at a similar Mach number).
It is interesting to note that the low-density tail of the  $\gamma>1$ PDF may flatten  further with resolution beyond $2048^{3}$ (this is seen in our model if a wider range of scales is kept):
 even at low velocities, very low density regions remain supersonic with strong shocks that cause large
density contrasts (see also \citealt{Federrath:2015gc} Fig.~4).

Finally, we note that similar ideas 
can be applied to MHD turbulence. A simple method, used in \citet{Padoan:2011jd,Molina:2012iv,Federrath:2015gc}
for the density variance-Mach number relation, is again  to consider 
how the Mach number is altered by the influence of the magnetic field on the total pressure,
\begin{equation}
\mathcal{M}\sim c_{s}^{-1} \sim \frac{\rho}{p_{\mathrm{gas}}+p_{\mathrm{mag}}} \sim c_{s0}^{-1}(1+ \beta^{-1})^{-1/2},\label{eq:B M hypothesis}\end{equation}
where $\beta = p_{\mathrm{gas}}/p_{\mathrm{mag}}$ is the ratio of thermal 
to magnetic pressure and $c_{s0}$ is the sound speed without the magnetic field. 
To apply this form of $\mathcal{M}$ to the density PDF model, we need a prescription for how $B$ changes with $\rho$. While
this 
remains uncertain, it unequivocally depends on turbulence parameters (e.g., Alfv\'en-Mach number and $\beta$; see \citealt{Lithwick:2001iv,2003MNRAS.345..325C,Burkhart:2009dm,2009MNRAS.398.1082B}). As an example, taking $B\sim \rho^{1/2}$ \citep{1999ApJ...520..706C,2009MNRAS.398.1082B,Molina:2012iv} we obtain
 a PDF of exactly the same functional form as the isothermal PDF, with shock sizes reduced by $1+\beta^{-1}$ (this
 is effectively identical to the model of  \citealt{Molina:2012iv}). In contrast,  a scaling
 $B \sim \rho^{\chi}$ with $\chi<1/2$ acts to decrease the PDF at
low densities, which is indeed seen in simulations \citep{Molina:2012iv}. A similar effect would be seen if $B(\rho)$ became constant 
below some density threshold\footnote{This form is suggested by the observations of \citet{2010ApJ...725..466C}, who report a
lower density bound below which the density and magnetic field are uncorrelated. It is also 
expected on physical grounds because the turbulence will become Alfv\'enic in character ($v_{A}\propto B/\rho^{1/2}>v_{l}$)
below some density \citep{Lithwick:2001iv,2003MNRAS.345..325C}.} (i.e., if the relation between $B$ and $\rho$ was not 
a power law, but a more sudden change).  
An interesting consequence of this is that the increased log-normality observed 
in MHD turbulence is probably not due to increased Gaussianity in the underlying turbulence. Instead, we may
be seeing suppression of the low-density tail of a compound-log-Poisson distribution,
which causes the PDF to appear log-normal even though the underlying 
turbulence could have similar intermittency properties [this this is the same effect as in  non-isothermal turbulence with $\gamma<1$; see Fig.~\ref{fig:non iso}(c)].
While there are many interesting (and astrophysically relevant) issues to explore here, 
we postpone such studies to future work due to the  uncertainties regarding the scaling 
of $B$ with $\rho$.

%%%%%%%%%%%%%%%%%%%%%%%%%%%%%%
\section{Discussion and conclusions}

In this paper, we propose a simple phenomenological model to describe the distribution of 
density in supersonic turbulence. Given the turbulent Mach number $\mathcal{M}$ and two free parameters ($\kappa$ and $\xi$) that relate to the physical properties of shocks, the model predicts the PDF of the 
density averaged over scale $l$, $\mathcal{P}_l(\ln \rho)$ [Eq.~\eqref{eq:model}]. Since  $\mathcal{P}_l(\ln \rho)$ 
completely specifies the statistics of the density field, the model predicts all relevant statistic quantities of the density field: the density variance--Mach number relation, the density PDF and intermittency, power spectra, and structure functions. We see reasonable agreement between  model predictions, results from previous literature, and our own set of simulations. The model is also straightforwardly extendable to more complex gas physics (e.g., varied gas equations of state, or magnetohydrodynamics) and shows decent agreement to recent simulations of nonisothermal turbulence \citep{Federrath:2015gc}.

The main predictions and results are summarized as follows:
\begin{itemize}
\item{The gas density, averaged across scale $l$, is distributed according to the PDF suggested in \citetalias{Hopkins:2013hn}; see Eq.~\eqref{eq:Phils}. The intermittency parameter $T$ controls 
the deviation from log-normality ($T=0$ describes a log-normal distribution). This form of the PDF matches numerical measurements very well across many orders of magnitude (see Fig.~\ref{fig:PDF and T}). The intermittency and variance  of the density PDF change with scale and Mach number.}
\item The density is arranged into a random collection of shocks across all scales (above the scale at which the velocity becomes subsonic). The physical size of the shocks and their relative density contrast are controlled by the model parameter $\kappa \sim r_{\mathrm{shock}}/l_{\mathrm{sonic}} $, which sets $T$ in the density PDF. Larger and higher-density shocks create a density distribution that is more intermittent.
\item Mathematically, the density is constructed via a compound-log-Poisson process. The size of each individual event (shock) is distributed according to an exponential distribution (see Fig.~\ref{fig:stepPDF}).
\item The number of shocks encountered across some range in scales is set by the maximum density that is possible if the gas is compressed in $\xi$ dimensions, where $\xi$ is effectively a model parameter (its maximum is $\xi=3$). Empirically, we find that $\xi\approx 1.3 \rightarrow 1.5 $ gives a reasonable fit to data, but given the significant scatter in previous results and between numerical methods (see Fig.~\ref{fig:dens variance}) the estimate is quite approximate.
\item The parameter $\kappa$ differs between compressibly and solenoidally forced turbulence, because the shocks are more intense with  compressive forcing \citep{Federrath:2013gu}. Based on numerical PDFs (see also App.~\ref{app:shock sizes} for a more direct measurement) $\kappa$ ranges from $\sim 0.2$ to $\sim 0.5$ as the compressive fraction is increased (but could also be lower in some cases; see \citealt{Pan:2015wk}).
\item The density variance--Mach number relation is similar to the standard result $S_{\ln \rho}\approx \log (1+ b^2 \mathcal{M}^2)$ for values of $\kappa$ that match the observed intermittency.
\item The model predicts a density power spectrum for $\mathcal{M}\gg 1$ between $\sim k^{-1}$ and $\sim k^0$, depending on $\kappa$ (and $\xi$)  [see Eq.~\eqref{eq:power law rho}], thus predicting a spectrum that is directly related to the intermittency. The power spectrum does not approach $k^0$ in the $\mathcal{M}\rightarrow\infty$ limit.
\item We neglect the influence of subsonic motions on the density PDF, since these are 
negligible at high $\mathcal{M}$ and stem from different physical processes. However, subsonic contributions to the PDF can reduce the intermittency (see Fig.~\ref{fig:PDF scales 1.3}) and may be responsible for some of the scatter seen across numerical results (see Fig.~\ref{fig:PDF and T}).
\item Extensions to the assumption of isothermal neutral gas may be included by considering the sound speed, and thus local shock size, to be a function of local gas density (as in \citealt{Passot:1998cr}). This leads to density PDFs that agree well with those observed in simulations with a nonisothermal equation of state (see Fig.~\ref{fig:non iso}).
\end{itemize}

More generally---particularly considering the very significant scatter between simulations reported in previous literature
(see, e.g.,  Figs.~\ref{fig:PDF and T} and \ref{fig:dens variance})---the model can be seen as a framework for understanding  density statistics and intermittency, describing how different statistical measures might be compared to provide interesting information about the underlying structures. This is particularly true in the moderate-$\mathcal{M}$ regime, which is most relevant physically and easiest to study numerically, but is not blessed with an true inertial range (due to the proximity of $l_{\mathrm{sonic}}$ to the scales of interest). For  example, as mentioned throughout the text, our simple model for the shock size $r_{\mathrm{shock}}\sim \kappa l_{\mathrm{sonic}}$, 
probably underestimates  the increase in $T$ with $\mathcal{M}$, which could stem from 
the mass-fraction contained in shocks increasing somewhat from $\mathcal{M}\sim 1$ to $\mathcal{M} \gg 1$ (i.e., $\kappa$ increasing with $\mathcal{M}$). 
In this vein, it is prudent to carry out further  tests of the model using higher-resolution numerical simulations (ideally with differing numerical methods), in particular, direct measurements of $\mathcal{P}_l(\ln\rho)$ (as in Fig.~\ref{fig:PDF scales 1.3}) across a range of $\mathcal{M}$. 

Of course, real turbulence in the ISM  involves a wide variety of other physical effects, which could strongly modify the ideal isothermal behavior discussed through most of this work.
For example, magnetic fields, self-gravity, dust, and more complicated radiation physics could all play key roles in some cases, and the ISM can hardly be considered  a homogeneous medium. Given the physical motivations behind various choices in the model, some of these features can be included in extended versions of the model (see Sec.~\ref{sec: Extensions}), albeit heuristically. Such extensions could be interesting to study in future work and potentially important for making astrophysically relevant predictions. 
The model could also form the basis for a description of other physical effects that are strongly influenced by turbulence. Star formation  \citep{Hopkins:2013jf,Hennebelle:2013fj} is an obvious example for 
such applications, but there are also a  variety  other possibilities; for instance, the dynamics of dust grains, which are key in controlling  the distribution of metals in the ISM \citep{draine:2003.dust.review} and strongly affected by  turbulence \citep{Hopkins:2015vc,Lee:2017md}.

\vspace{-0.7cm}
\acknowledgments 
JS was funded in part by the Gordon and Betty Moore Foundation through Grant GBMF5076 to Lars Bildsten, Eliot Quataert and E. Sterl Phinney. Support for PFH was provided by NASA ATP Grant NNX14AH35G \&\ NSF Collaborative Research Grant \#1411920 and CAREER grant \#1455342.  Numerical calculations were run on Caltech cluster ``Zwicky'' (NSF MRI award \#PHY-0960291) \&\ XSEDE allocation TG-AST130039 supported by the NSF.

\vspace{0.7cm}

\appendix
\section{Derivation of the density structure functions and power spectra from $\mathcal{P}_{l}(\ln\rho)$}\label{app:PS derivation}

In this appendix we illustrate how to derive structure functions and power
spectra from the scale variation variation of the PDF $\mathcal{P}_{l}(\ln\rho)$.
This derivation is completely general and could apply to any statistical field for which one had $\mathcal{P}_{l}$.

Consider the variable $s$, denoting its average over scale $l $ as $s_{l}$. We take $s_{l}$ distributed  according to the PDF $s_{l}\sim \mathcal{P}_{l}(s_{l})$,  with mean $\bar{s}_{l} = \int s_{l} \mathcal{P}_{l}(s_{l}) ds_{l}  $ and variance $S_{l} = \int s_{l}^{2} \mathcal{P}_{l}(s_{l}) ds_{l} - \bar{s}_{l}^{2}$. 
The first observation is that the isotropic autocorrelation of $s_{l'}$ 
\begin{equation}
R_{s_{l'}}(l) = R_{s_{l'}}(|\bm{l}|) = \langle s_{l'}(\bm{x}+\bm{l})s_{l'}(\bm{x})\rangle.
 \end{equation}
is the same as $R_{s_{l''}}(l)$, so long as $l''<l$ and $l'<l$. We also note that if $l'>l$, 
 \begin{equation}
R_{s_{l'}}(l)= S_{l} + \bar{s}_{l}^{2},
\end{equation}
which follows because $s_{l'}$ is constant on scales less than $l'$. This implies 
\begin{equation}
R_{s}(l)= S_{l} + \bar{s}_{l}^{2},\end{equation}
relating direct measurements of $s$ to $\mathcal{P}_{l}$.

The second-order structure function is now easily calculated as
\begin{equation}
\langle \Delta s^{2} \rangle = \langle  [s(\bm{x}+\bm{l})-s(\bm{x})]^{2}\rangle =2 \langle s^{2} \rangle - 2 (S_{l} + \bar{s}_{l}^{2}) \label{app:eq: SF scaling full}.\end{equation}
However,  the nonlocality  of Eq.~\eqref{app:eq: SF scaling full} (it depends on the smallest scales through $\langle s^{2}\rangle$) is inconvenient, and it is more helpful to  consider its derivative, 
\begin{equation}
\frac{1}{2}\frac{d\langle \Delta s^{2} \rangle}{dl}  = - \frac{d}{dl}( S_{l}+s_{l}^{2}).\label{app:eq: struct func}\end{equation}
Equation~\eqref{app:eq: struct func} is more useful than Eq.~\eqref{app:eq: SF scaling full} because it lacks dependence on the smallest or largest scales in the system.

The 3-D power spectrum $\Phi_{s} (\bm{k})$ is related to the autocorrelation through the standard Fourier 
transform.  However, since we are interested in the 1-D spectrum $4\pi k^{2} \phi_{s}(k) = \Phi_{s}(\bm{k})$,
the relation is instead \citep{Davidson:2015wl}
\begin{equation}
R_{s}(l) = \int_{0}^{\infty}\phi_{s}(k) \mathrm{sinc}(kl) dk,\label{eq:3D FT relation}
\end{equation}
where $\mathrm{sinc}(x) = x^{-1}\sin(x)$. Approximating $\mathrm{sinc}(k l) \approx \Pi (k l/2\pi)$, where $\Pi(x)$ is the tophat function ($1$ for $-1<x<1$, $0$ otherwise), we see that the transform \eqref{eq:3D FT relation} is related to a filtering operation
\begin{equation}
R_{s}(l) \approx \int_{2 \pi/l}^{\infty} \phi_{s}(k) dk, \end{equation}
or
\begin{equation}
\frac{d \langle \Delta s^{2}\rangle}{dl} \approx \frac{2\pi}{l^{2}}\phi_{s}\left(\frac{2\pi}{l}\right).\label{app:eq: SF to PS}
\end{equation}
Neglecting numerical constants (we are only interested in the $k$-scaling of the power spectrum), we obtain  
\begin{equation}
\phi_{s}(k) \sim \frac{d}{dk}( S_{l}+s_{l}^{2}),\end{equation}
which fits with the intuition that the power spectrum should encode the change in the variance of $s$ with scale.

 \section{Shock sizes}\label{app:shock sizes}
 In this appendix, we explicitly test the  assumptions about shock width $r_{\mathrm{shock}}$ that went into 
 deriving $T$.
In particular, the scaling $r_{\mathrm{shock}}\sim \kappa l_{\mathrm{sonic}}$ was important for relating
the mathematical properties of the model to physical characteristics of the turbulent density field.  To test this, we measure $r_{\mathrm{shock}}$ from simulation and compare this to $\kappa l_{\mathrm{sonic}}$. While 
not technically a test of the model, this  is important to  
verifying that $\kappa$, as measured from the intermittency of the PDF through $T$,
is broadly consistent with the true width of shock structures. In other words, having seen in  Sec.~\ref{sec:Numerics} 
that the model gives decent predictions of turbulent statistics, is our physical 
interpretation $\kappa\sim r_{\mathrm{shock}}/l_{\mathrm{sonic}}$ consistent with the properties of shocks seen in simulations? 
\begin{figure}
\begin{center}
\includegraphics[width=1.0\columnwidth]{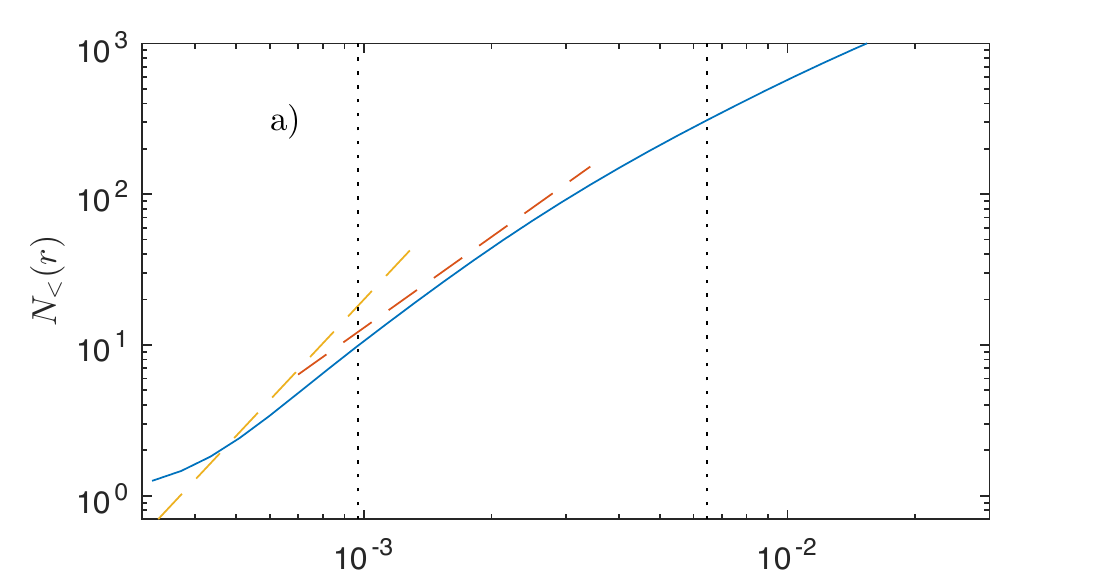}
\includegraphics[width=1.0\columnwidth]{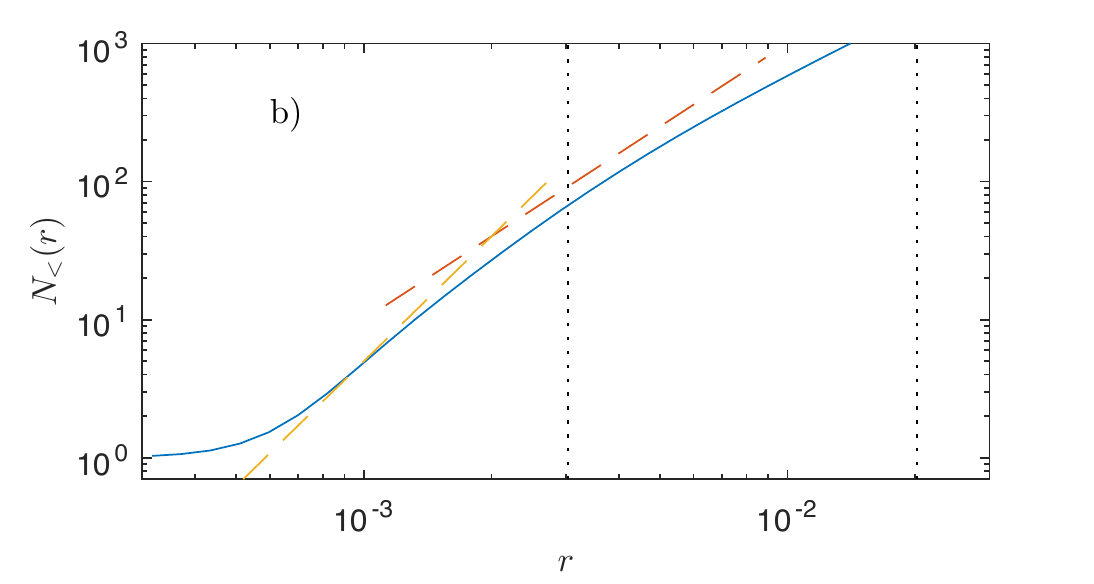}
\caption{Average of $N_{<}(r) $ over snapshots and 10000 randomly chosen center cells
for (a) $\mathcal{M}\approx 12$ and (b) $\mathcal{M}\approx 7$. The dashed red and yellow
lines show $r^{2}$ and $r^{3}$ scalings respectively. 
The vertical dashed lines show the inferred sonic scale $l_{\mathrm{sonic}}\sim \mathcal{M}^{-2}$ (right) and 
$\kappa l_{\mathrm{sonic}}$ (left) with $\kappa = 0.15$. Although of narrow extent due to the flattening of $N_{<}(r)$ to $N_{<}(r)=1$ for $r<r_{\mathrm{cell}}$, there is a region at $r$ where 
the scaling is substantially steeper than $r^{2}$. 
The transition to $N_{<}(r)\sim r^{2}$ scaling occurs significantly below $l_{\mathrm{sonic}}$ but in 
rough agreement with $\kappa l_{\mathrm{sonic}}$  for both $\mathcal{M}\approx 12$ and $\mathcal{M}\approx 7$.
}
\label{fig:shock sizes}
\end{center}
\end{figure}

The Lagrangian nature of the GIZMO code implies that the density is directly 
proportional to the number density of cells. We thus use a simple  counting method 
to measure $r_{\mathrm{shock}}$. This involves defining all cells with $\rho > \rho_{\mathrm{max}}/10$
(where $\rho_{\mathrm{max}}$ is the maximum density across the current snapshot)
as being part of a ``shock,''  then counting the number of such cells within radius $r$, $N_{<}(r)$, of a randomly chosen center
cell. Ideally, if $r<r_{s}$, then $N_{<}(r)\sim r^{3}$, while if $r>r_{s}$ then $N_{<}(r)\sim r^{\upsilon}$, where 
$\upsilon\approx 2$ is the fractal dimension of the shock \citep{Federrath:2008ey}. 
We carry out this procedure for 10000 randomly chosen center
cells per simulation, averaging the results, then averaging these
results over time in the statistical steady state of the turbulence.
Of course, in the messiness of a true turbulent density field, the transition at $r\sim r_{s}$ will
 be relatively smooth, and it is difficult to unambiguously define $r_{\mathrm{shock}}$. In addition, 
  at very small scales $N_{<}(r)$  is adversely affected by the finite number 
of cells, since $N_{<}(r)\rightarrow 1$ as $r < r_{\mathrm{cell}}$  (where $r_{\mathrm{cell}}$ is the 
size of a cell in the shocked region).
 Note that this method assumes an approximately constant 
distribution of cells inside the shock, which appears to be the case based on examination of
2-D density field slices.

Results are shown in Fig.~\ref{fig:shock sizes} for the $\mathcal{M}\approx 12$ and $\mathcal{M}\approx 7$ simulations from above.
In both cases there is a clear flattening  to $N_{<}(r)\sim 2$ (the further 
flattening at higher $r$ is probably related to the finite extent of the nearly 2-D high density regions). The key point 
is that this flattening occurs well below  $r\sim l_{\mathrm{sonic}}=\mathcal{M}^{-2}$, in approximate agreement 
with the estimate $r\sim \kappa l_{\mathrm{sonic}}$ with $\kappa \sim 0.2$. Thus, 
while it is difficult to accurately measure the shock width, our
hypothesis that $r_{\mathrm{shock}}\sim \kappa l_{\mathrm{sonic}}$ is consistent 
with the data, while $r_{\mathrm{shock}}\sim l_{\mathrm{sonic}}$ provides a significant overestimate of the width. 
Future simulations at higher resolution may allow for a more accurate determination of these properties,
by more accurately capturing the transition from supersonic to subsonic motions and allowing measurements over a wider range of $\mathcal{M}$.
 Unfortunately, even with the Lagrangian numerical method, the shock size at $\mathcal{M}\approx 40$ is 
still too small to see a transition from $N_{<}(r)\sim r^{2}$ to  $N_{<}(r)\sim r^{3}$ (i.e., $r_{\mathrm{cell}}<r_{\mathrm{shock}}$), so we do not plot this here.

%\bibliographystyle{apj}
%\bibliography{fullbib}

%%%%%%%%%%%%%%%%%%%%%%%%%%%%%%%%%
%%%%%%%%%%%%%%%%%%%%%%%%%%%%%%%%%
%%%%%%%%%%%%%%%%%%%%%%%%%%%%%%%%%
%%%%%%%%%%%%%%%%%%%%%%%%%%%%%%%%%

\end{document}